\documentclass[%
 aip,
 jcp,
 amsmath,
 amssymb,
 preprint,
]{revtex4-2}

\usepackage{graphicx}
\usepackage{dcolumn}
\usepackage{bm}

\usepackage[utf8]{inputenc}
\usepackage[T1]{fontenc}
\usepackage{mathptmx}

\usepackage[acronym]{glossaries}
\newacronym{DFT}{DFT}{density functional theory}
\newacronym{TDDFT}{TDDFT}{time-dependent density functional theory}
\newacronym{SnPc}{\textbf{SnPc}}{tin(II) phthalocyanine}
\newacronym{CE}{\textbf{SnPc-Ag}}{\textbf{SnPc-Ag}}
\newacronym{FE}{\textbf{SnPc}$\pm$E}{\textbf{SnPc}$\pm$E}
\newacronym{FULL}{\textbf{SnPc-Ag}$\pm$E}{\textbf{SnPc-Ag}$\pm$E}
\newacronym{IMDHOM}{IMDHOM}{independent-mode displaced harmonic oscillator model}
\newacronym{LMCT}{LMCT}{ligand-to-metal charge transfer}
\newacronym{MLCT}{MLCT}{metal-to-ligand charge transfer}
\newacronym{MC}{MC}{metal-centered}
\newacronym{SERS}{SERS}{surface-enhanced Raman spectroscopy}
\newacronym{TERS}{TERS}{tip-enhanced Raman spectroscopy}

\usepackage{lmodern}
\usepackage[ngerman,english]{babel}
\usepackage{amsmath}
\usepackage{amsfonts}
\usepackage{amssymb}
\usepackage{nicefrac}
\usepackage[
			]{siunitx}
\DeclareSIUnit{\arbu}{arb.u.}
\DeclareSIUnit{\au}{a.u.}

\usepackage{color}
\usepackage{soul}

\begin{document}


\title{A full quantum mechanical approach assessing the chemical and electromagnetic effect in TERS}


\author{K. Fiederling}
\affiliation{Institute of Physical Chemistry and Abbe Center of Photonics, Friedrich Schiller University Jena, Helmholtzweg 4, 07743 Jena, Germany}
\author{M. Abasifard}
\affiliation{Institute of Applied Physics, Friedrich Schiller University Jena,  Albert-Einstein-Str. 15, 07745 Jena, Germany}
\affiliation{Institute of Solid State Physics, Friedrich Schiller University Jena, Helmholtzweg 3, 07743 Jena, Germany}
\author{M. Richter}
\affiliation{Present address: DS Deutschland GmbH, Am Kabellager 11-13, 51063 Cologne, Germany}
\author{V. Deckert}
\affiliation{Institute of Physical Chemistry and Abbe Center of Photonics, Friedrich Schiller University Jena, Helmholtzweg 4, 07743 Jena, Germany}
\affiliation{Leibniz Institute of Photonic Technology (IPHT), Albert-Einstein-Str. 9, 07743 Jena, Germany}
\author{S. Kupfer}
\email{stephan.kupfer@uni-jena.de}
\author{S. Gräfe}
\email{s.graefe@uni-jena.de}
\affiliation{Institute of Physical Chemistry and Abbe Center of Photonics, Friedrich Schiller University Jena, Helmholtzweg 4, 07743 Jena, Germany}

\date{\today}

\begin{abstract}

Tip-enhanced Raman spectroscopy (TERS) is a valuable method for surface analysis with nanometer to angstrom-scale resolution, however, the accurate simulation of particular TERS signals remains a computational challenge.
We present a unique approach to this challenge by combining the two main contributors to plasmon-enhanced Raman spectroscopy and to the high resolution in TERS in particular, the electromagnetic and the chemical effect, into one quantum mechanical simulation. 
The electromagnetic effect describes the sample's interaction with the strong, highly localized and inhomogeneous electric fields associated with the plasmonic tip, and is typically the thematic focus for most mechanistic studies. 
On the other hand, the chemical effect covers the different responses to the extremely close-range and highly position-sensitive chemical interaction between the apex tip atom(s) and the sample, and, as we could show in previous works, plays an often underestimated role. 
Starting from a (time-dependent) density functional theory description of the chemical model system, comprised of a \acrlong{SnPc} sample molecule and a single silver atom as tip, 
we introduce the electromagnetic effect through a series of point charges that recreate the electric field in the vicinity of the plasmonic Ag nanoparticle. 
By scanning the tip over the molecule along a 3D grid, we can investigate the system's Raman response on each position for non-resonant and resonant illumination. 
Simulating both effects on their own already hints at the achievable signal enhancement and resolution, but the combination of both creates even stronger evidence that TERS is capable of resolving sub-molecular features.

\end{abstract}

\maketitle

\section{Introduction}

Since its first experimental demonstration in the year 2000, \acrlong{TERS}\cite{Stockle_ChemicalPhysicsLetters_2000, Hayazawa_OpticsCommunications_2000, Anderson_Appl.Phys.Lett._2000} has become an immensely useful tool for the study of surface-bound molecules. 
Under certain conditions, even sub-molecular features can be resolved\cite{Zhang_Nature_2013, Fang_Phys.Chem.Chem.Phys._2014,Klingsporn_J.Am.Chem.Soc._2014,Sun_Adv.Opt.Mater._2014,He_J.Am.Chem.Soc._2019,Jiang_Nat.Nanotechnol._2015,Lee_Nature_2019,Richard-Lacroix_Chem.Soc.Rev._2017},  
which was controversially discussed as this is way beyond the limit of conventional light microscopy. 
Similar to surface enhanced Raman spectroscopy (SERS)\cite{Fleischmann_ChemicalPhysicsLetters_1974,Albrecht_J.Am.Chem.Soc._1977,Jeanmaire_JournalofElectroanalyticalChemistryandInterfacialElectrochemistry_1977,Langer_ACSNano_2020}, 
the basis of TERS is the surface plasmon generated by irradiating a nanoparticle (or nano-structured tip) close to the analyte.\cite{Jiang_NanoLett._2012,Pettinger_Phys.Rev.Lett._2004}

The exact nature of the underlying mechanisms that enable such a high resultion are still under investigation, but are broadly categorized into two contributions\cite{Jeanmaire_JournalofElectroanalyticalChemistryandInterfacialElectrochemistry_1977}.
The first of these is the electromagnetic effect\cite{Corni_ChemicalPhysicsLetters_2001, Corni_J.Chem.Phys._2001a, Chulhai_J.Phys.Chem.C_2013, Chiang_NanoLett._2015, Zhang_J.Phys.Chem.C_2015, Morton_Chem.Rev._2011, Thomas_J.RamanSpectrosc._2013}, 
describing how the highly inhomogeneous electromagnetic fields from the plasmonic tip interact with the sample and enhance the Raman response, 
especially in a so-called pico-cavity between the plasmonic tip and a possible metal substrate.\cite{Zhang_Phys.Rev.B_2014, Benz_Science_2016, Trautmann_Nanoscale_2016, Chen_Nanoscale_2018, Urbieta_ACSNano_2018, Baumberg_Nat.Mater._2019}
The other one being the chemical effect\cite{Zhao_J.Am.Chem.Soc._2006, Jensen_J.Phys.Chem.C_2007, Jensen_Chem.Soc.Rev._2008, Liu_SpectrochimicaActaPartA:MolecularandBiomolecularSpectroscopy_2009, Valley_J.Phys.Chem.Lett._2013, Latorre_Phys.Chem.Chem.Phys._2015, Morton_Chem.Rev._2011},
describing how the molecule's electronic structure changes while in proximity to the plasmonic particle. 
In TERS, this is especially relevant, since the scanning tip can be moved very precisely to different areas of the sample molecule, exhibiting site-specific responses to the changeing chemical environment.

Several studies focus heavily on the plasmon in different nanoparticle configurations and therefore mainly try to understand the electromagnetic effect, as published by the groups of Aizpurua\cite{Zhang_Nature_2013,Barbry_NanoLett._2015,Benz_Science_2016,Schmidt_ACSNano_2016,Langer_ACSNano_2020, Zhang_JRamanSpectrosc_2021}, 
Jensen\cite{Payton_J.Chem.Phys._2012,Payton_Acc.Chem.Res._2014,Hu_J.Chem.TheoryComput._2016,Liu_ACSNano_2017,Chen_Nanoscale_2018,Chen_Nat.Commun._2019}, 
or Schatz\cite{Hao_J.Chem.Phys._2003,Zou_J.Chem.Phys._2004,Gieseking_J.Phys.Chem.A_2016,Ding_J.Phys.Chem.C_2018, Kluender_J.Phys.Chem.C_2021}.
In particular, the discrete interaction model/quantum mechanics (DIM/QM) method \cite{Morton_J.Chem.Phys._2010, Morton_J.Chem.Phys._2011, Payton_J.Chem.Phys._2012} operates on a  similar quantum chemical level of theory, 
and includes an atomistic model for the nanoparticle, however, no explicit metal atoms are present in the quantum chemical simulations.
Most approaches typically assume a kind of 'medium-range' regime, where the plasmonic particle is sufficiently close to the sample to meaningfully enhance the Raman signal, but distant enough to neglect the chemical interactions.
Since the electromagnetic effect is already relatively well-described with classical electrodynamics\cite{Zhao_Acc.Chem.Res._2008}, we follow a strategy that builds a model system from the chemical point of view.

Our quantum mechanical approach starts from a high-level simulation of the sample molecule and incorporates the plasmonic tip into those simulations as accurately as possible. 
To properly account for the chemical effect, a small part of the nanoparticle is fully incorporated into the quantum chemical simulations, in the form of the apex atom. 
Extending this previously established\cite{Latorre_Nanoscale_2016, Fiederling_Nanoscale_2020,  Fiederling_J.Chem.Phys._2021, Rodriguez_ACSPhotonics_2021}  model, the electric field in vicinity to the plasmonic nanoparticle is added to these simulations
through a series of point charges.
This tip model (apex atom and field-creating point charges) is then scanned over the immobilized sample molecule, \acrfull{SnPc}, 
and allows the study of short-range chemical interaction between tip and molecule, as well as the influence of the much larger plasmonic nanoparticle on the whole system. 
Additionally, since both parts of the tip model are separable, we can evaluate and quantify the system's behaviour if only one of the two effects is active and therefore study how they influence each other.

The results of this new, purely quantum mechanical approach are presented in the following contribution. 
In the first part, the energy of the incident radiation is assumed to be far from any electronic excitations of the sample molecule, i.e. non-resonant conditions. 
Here, the chemical effect (no external point charges) and the electromagnetic effect (no explicit tip atom) are investigated separately first, and then how the full model system behaves, with respect to the two contributions.
The second part focuses on resonant conditions, and how the included electric fields influence the molecular properties that are relevant for the system's resonant Raman response under the IMDHO model.
While the same quantum chemical system (\textbf{SnPc-Ag}) has been investigated before\cite{Fiederling_Nanoscale_2020}, the addition of the external electric fields proves to be a valuable extension of the model and provides further insight into the mechanism governing TERS.

\section{Computational Details}

\subsection{Finite Element Simulations}

The RF module of Comsol Multiphysics\textregistered{} 5.0\cite{comsol} 
was used to solve Maxwell’s equations in a finite element setup in order to obtain the field distribution in the vicinity of the silver tip. 
Here, the underlying equation that governs the electric field is: 
\begin{equation}
\nabla^2 E - k_0^2  \epsilon_\mathrm{rc} E = 0
\end{equation}
with the permittivity $\epsilon_\mathrm{rc}$ and the relative permeability $\mu_r$ set to 1 
and the wave number of the free space $k_0$ depending on the speed of light $c_0$:
\begin{equation}
 k_0 = \omega \sqrt{\epsilon_0 \mu_0} = \frac{\omega }{c_0}
\end{equation}

The following computational setup is based on the procedure introduced by Trautmann et al.\cite{Trautmann_Nanoscale_2016}
The Simulations were performed in a cubic system of size 700x700x\SI{700}{\cubic\nano\meter} 
with periodic boundary conditions (Floquet periodicity) at the side walls ($x$- and $y$-direction) 
and top and bottom walls ($z$-direction) defined as periodic port boundaries. 
For modelling the field distribution near the tip, a spherical nanostructure (radius \SI{100}{\angstrom}) above a flat surface, acting as substrate 
(glass, $n_\mathrm{glass}=1.520$), was created. 
A substructure was modelled as solid of revolution from a Gaussian profile resembling a single atom that protrudes from the otherwise smooth surface, 
creating an atomic scale plasmonic substructure. 
The Gaussian profile was defined as:
\begin{equation}
 f(x, \chi, \zeta) = \frac{\zeta}{\sqrt{2 \pi}} e^{- \frac{x^2}{2 \chi}}
\end{equation}
with $\chi= \SI{2.11}{\angstrom}$ determining the width of the Gaussian profile 
and $\zeta = \SI{8.5}{\angstrom}$ resembling the protrusion for a silver atom with an atomic radius of \SI{1.65}{\angstrom}. 
The gap between the substrate and the lowest point of the nanoparticle was set to $z_g = \SI{10}{\angstrom}$. 
Material settings for the silver nanoparticle were taken from Ref. \cite{Johnson_Phys.Rev.B_1972}. 

The incident field, with an electric field strength of \SI{5.142e8}{\volt\per\meter} (\SI{e-3}{\au}),\cite{Sakko_JPhysCondensMatter_2014} 
entered the system from the bottom port with an angle of $\theta = \SI{60}{\degree}$, where the incident power was obtained as:
\begin{equation}
 P = \frac{1}{2} E_0^2 n_\mathrm{air} c_0 \epsilon_0 \omega^2 \cos{\theta}
\end{equation}
with $c_0, \epsilon_0, n_\mathrm{air} = 1 $ being the speed of light, vacuum permittivity, and air refractive index, respectively. 
The minimal tetrahedral mesh size was set to \SI{0.9}{\angstrom} in a \SI{100}{\nano\meter} sphere around the nanoparticle 
and \SI{5}{\nano\meter} otherwise to reduce the computational cost.

\begin{figure*}
 \includegraphics{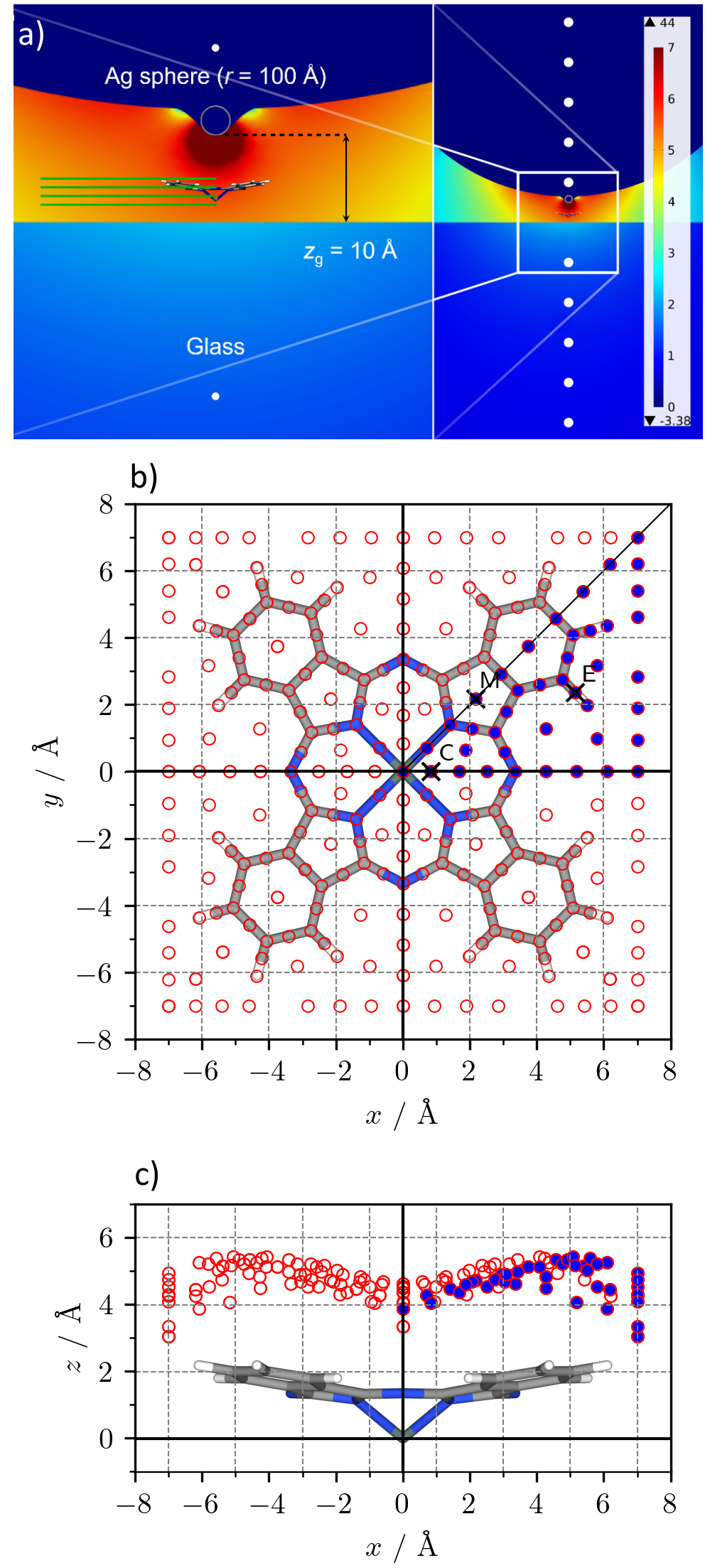}
 \caption{\label{fig:com_det} a) Setup for the COMSOL simulation with field-creating point charges (white), calibration points shown in green.
 b) and c): Input orientation for the QC simulations along $z$- and $y$-axis, respectively. Unique Ag-positions (blue dots) that are mirrored/rotated to yield all possible positions (red circles) above the \textbf{SnPc} molecule. Examplarily discussed positions 'center', 'middle', and 'edge' marked as C, M, E respectively.}
\end{figure*}

The electric field distribution obtained by the finite element calculations was then mimicked by ten point charges 
- five above and five below the glass substrate at 
$z =  \pm2, \pm3, \pm4, \pm5, \mathrm{ and} \pm\SI{6}{\nano\meter}$,
respectively, see Fig.~\ref{fig:com_det}a)) (white dots). 
Two sets of these ten point charges were obtained to model the oscillating electric field at opposing field orientations: 
the positive (crest) and negative (trough) extreme values. 
Each set of point charges was introduced in the subsequent quantum chemical calculations employing Gaussian16\cite{g16}, respectively, 
to mimic the field distribution in the vicinity of \textbf{SnPc-Ag}, \textit{i.e.}, 
at four layers of calibration points covering the investigated molecule and located 
2 (90 grid points), 3 (129 grid points), 4 (140 grid points), and \SI{5}{\angstrom} (151 grid points) above the glass substrate, see Fig.~\ref{fig:com_det}a)) (green dots). 
The electric field at these calibration points, modelled by these point charges, 
differs from the electric field obtained by the preliminary finite element simulations merely by \SI{1.5}{\percent} (mean absolute deviation), 
while a maximum deviation of less than \SI{4}{\percent} is obtained at the edge of the calibration layers (at $x,y = \approx \SI{15}{\angstrom}$).
The magnitude of the charges is scaled so that the field strength at the molecule's position is about \SI{3}{\giga\volt\per\meter}.

\subsection{Quantum Chemical Simulations}

The simulation of the full \acrshort{TERS} experiment follows our previously published
\cite{Latorre_Nanoscale_2016,Fiederling_J.Chem.Phys._2021, Fiederling_Nanoscale_2020} (TD)DFT-based
protocol, with the inclusion of the above mentioned point charges to complement the tip model.
All quantum chemical calculations were carried out with Gaussian16\cite{g16} on the \acrshort{DFT} or \acrshort{TDDFT} level of theory, respectively, 
using the range-separated CAM-B3LYP\cite{Yanai_ChemicalPhysicsLetters_2004} functional 
and 6-311+G**\cite{Krishnan_J.Chem.Phys._1980, Clark_J.Comput.Chem._1983} triple-$\zeta$ basis set.
For the heavy atoms (Sn, Ag) the electronic core potentials MWB46 and MWB28 
and their respective basis sets\cite{Andrae_Theoret.Chim.Acta_1990} were used.
The D3 dispersion correction with Becke-Johnson dampening was employed for all calculations.\cite{Grimme_J.Comput.Chem._2011}
The added point charges follow the movement of the tip atom in the $xy$-plane, above and below the \textbf{SnPc} molecule.
All considered tip positions are shown in Fig.~\ref{fig:com_det}b) and c), the three positions that are discussed in detail are highlighted. 

To elucidate the interplay of  the chemical and the electromagnetic effect, in the \acrshort{FULL} system, two sub-systems were investigated as well. 
One without the field-creating point charges, \acrshort{CE}, 
where only the chemical interaction between the Ag atom and the sample molecule are observed,
and one without the Ag atom, \acrshort{FE}, that only shows the molecule's response to the electric fields.
The \acrshort{CE} system, and with it the chemical effect, was discussed  in-depth previously in \textcite{Fiederling_Nanoscale_2020}.

In line with previous studies, the detection of the Raman signal is assumed to be mostly in $z$-direction and therefore the intensity of each mode is dependent on the $zz$-component of the respective derivative of the (transition) polarizability tensor
\begin{equation}\label{eq:int}
I_l = \left( E_\mathrm{L} - E_l \right) ^4 \left| \frac{\partial \left( \alpha_{zz} \right)_{g0_l \rightarrow g1_l}}{\partial q_l} \right| ^2 \;.
\end{equation}

For the non-resonant case, the polarizability derivatives are calculated directly by the quantum chemistry package. 
In the resonant case, transition polarizability derivatives are required and are calculated using the \acrfull{IMDHOM}\cite{Guthmuller_J.Chem.Phys._2016}
\begin{equation}\label{eq:alpha}
\left(\alpha_{zz}\right)_{g0_l \rightarrow g1_l} = \sum_e \left(\mu_z \right)_e^2 \frac{\Delta_{e,l}}{\sqrt{2}} \left( \Phi_e (E_\mathrm{L}) - \Phi_e (E_\mathrm{L} - E_l )\right) \;.
\end{equation}
Here, $\Delta_{e,l}$ is the dimensionless displacement, 
 $E_{e,g}$ the vertical excitation energy from the ground-state $g$ to excited-state $e$ 
 and $\Gamma$ a damping factor describing homogeneous broadening (chosen as $\SI{3000}{\per\centi\meter} \widehat{=} \SI{0.372}{\electronvolt}$), 
while $E_\mathrm{L}$ represents the energy of the irradiating laser.
The function $\Phi_e$, neglecting Franck-Condon factors, is given by:
\begin{equation}
 \Phi_e (E_\mathrm{L}) = \frac{1}{E_{e,g} - E_\mathrm{L} -\mathrm{i} \Gamma} \;.
\end{equation}

\section{Results}

In this section, we present the simulated \acrshort{TERS} response of the full \acrshort{FULL} system and compare it to the responses of both sub-systems (\acrshort{CE} and \acrshort{FE}) as well as to the free \acrshort{SnPc} molecule. 
This allows us to separate the chemical (sub-system \acrshort{CE}, in-depth discussed in \citet{Fiederling_Nanoscale_2020}) from the electromagnetic effect (\acrshort{FE}), as modelled by the applied electric fields, and see how both interact with each other.
First, the non-resonant response is evaluated with incident radiation at \SI{1064}{\nano\meter}, assumed to be far away from any electronic excitations.
In the second part, only the resonant response to the incident light is investigated, with its wavelength chosen according tothe predicted electronic excitations.

\subsection{Non-Resonant}

\begin{figure*}
 \includegraphics{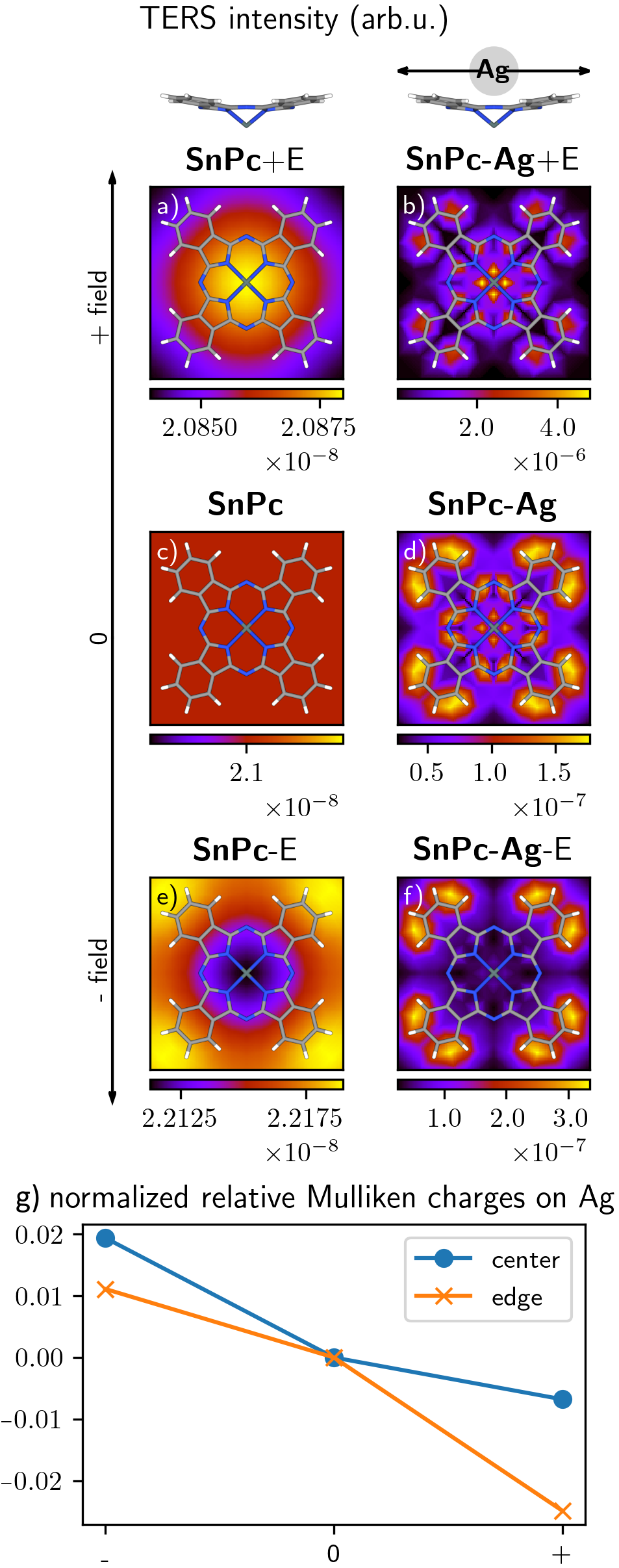}
 \caption{ \label{fig:non-res} a)-f): TERS Intensity maps (arb.u.). Middle row: no field, top/bottom row: +/- field, respectively. Left/right column: without/with explicit tip atom (Ag).  c) only shows one number, hence a flat color.
 g): Mulliken charges, normalized to 'middle' position with respective field, relative to the respective field-free case.}
\end{figure*}

Applying the non-homogeneous electric field from the COMSOL calculations to the \acrshort{SnPc} molecule 
without adding the Ag atom at the tip's apex barely affects the molecules \acrshort{TERS} response, as Fig.~\ref{fig:non-res} a) and e) shows.
The integrated intensity slightly increases for the negative field direction, while the positive direction slightly decreases the intensity. 
In both cases, the variation in intensity over the whole map is small, but exhibits one common feature: 
The central area of the molecule is brighter for the positive field direction and darker for the negative direction with respect to the map's edges.
Otherwise, there is no indication of any sub-molecular resolution.
A possible explanation could be an increased electron density on the very polarizable and slightly out-of-plane outer $\pi$-system in case of the negative field direction and the reverse effect for positive direction. 
The outermost hydrogen atoms are roughly \SI{2.2}{\angstrom} closer to the field-creating point charges above the tip than the central Sn atom, resulting in a \SI{3}{\percent} deviation in field strength in $z$-direction.

For the other sub-system with only the tip atom and no external field, the response is very different.
The overall intensity increases by roughly one order of magnitude and there are several equally bright, sub-molecular features.
Those features correspond  to three distinct regions over the molecule: 
the center directly around the Sn atom, the middle 'ring' on top of the alternating C-N atoms, and the edge above the C-H bonds.

Combining both sub-systems to form the full system with the Ag atom at the tip apex and the electric field (Fig.~\ref{fig:non-res} b) and f)), 
signal intensity generally rises and at least some of the sub-molecular structures from Fig.~\ref{fig:non-res} d) are still present. 
In case of the positive field direction, this increase in intensity is roughly one order of magnitude with regard to the chemical sub-system 
and two orders of magnitude with regard to the electrostatic sub-system. 
All three features from the field-free chemical sub-system are still present in the map,
but the 'edge' and 'middle' areas show less intensity than the bright 'center' structure.
This corresponds well with the Ag-free case (Fig.~\ref{fig:non-res} a), where the intensity in the center of the map is slightly higher.
For the negative field direction, only the 'edge' structure remains bright, while 'middle' and 'center' vanish into the background.
The maximum intensity increases barely by a factor of 2.
Similar to the positive case, this fits the result from the respective Ag-free sub-system, where the center of the map exhibits a slightly lower intensity.
In both cases, the resulting maps for the whole system reveal that both effects are interacting heavily with each other, 
boosting the overall degree of detail and intensity.

The different behaviour of the three features under opposite field directions can be explained by looking at the field-induced partial charge on the Ag atom.
For better visualization, the charges shown are relative to the charge at the 'middle' position in the respective sub-system, as the charge at this position is the least influenced by changing electric fields.
Additionally, the charges are normalized to the respective position's charge with no external electric field, leading to Fig.~\ref{fig:non-res}g).
Here, it is evident that the 'center' structure is far more sensitive to the positive field than the negative one, reflecting the bright and dark 'center' in the respective map.
Especially in the case of negative field direction,  charge and intensity are barely influenced.
On the 'edge' position, the behaviour is reversed: The charge is strongly affected by the negative field, resulting in a high intensity there.
 The positive field does not have an equally strong effect on the tip, and the map shows a less pronounced intensity on this position.

 \begin{figure*}
 \includegraphics[width=\textwidth]{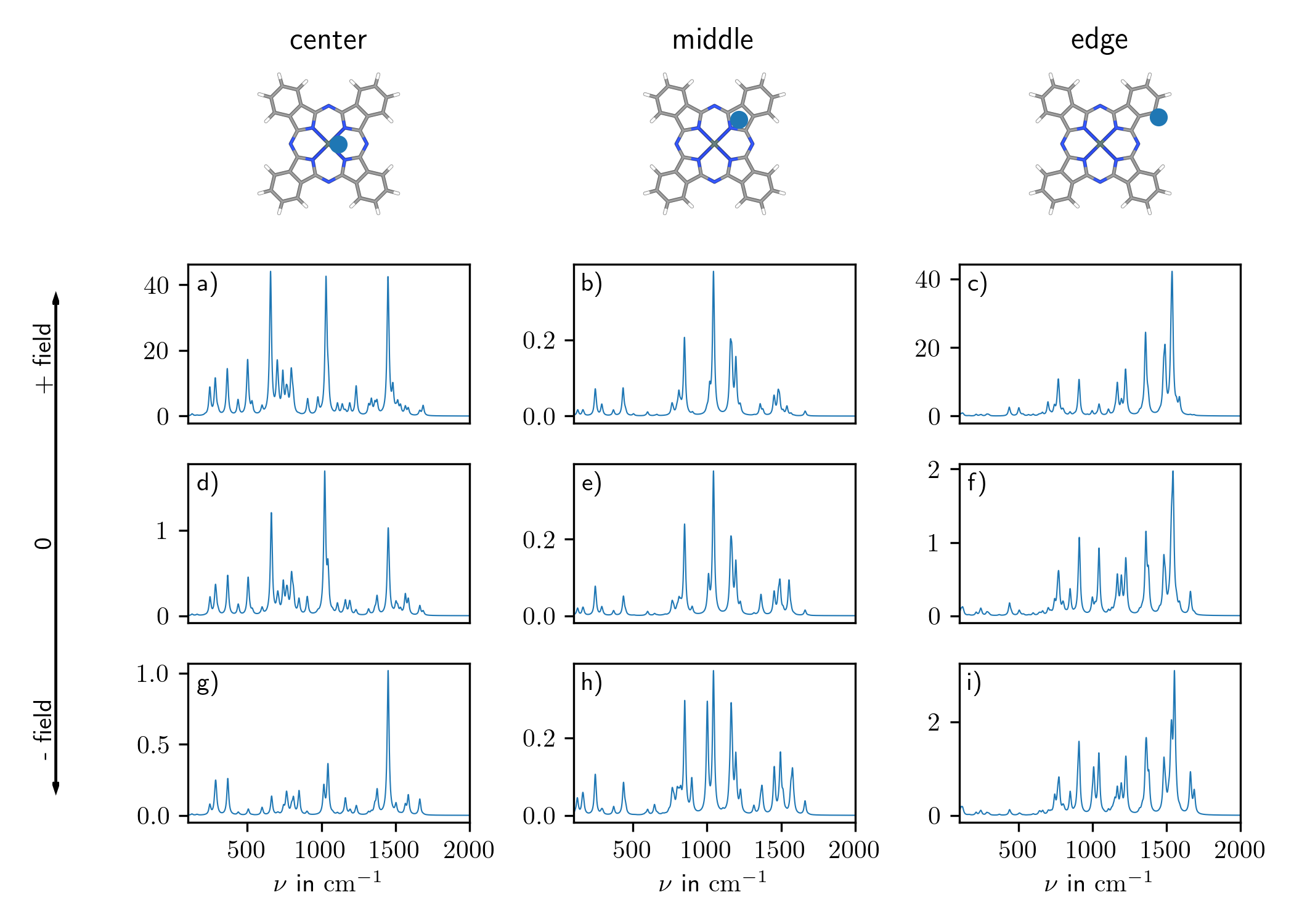}
 \caption{\label{fig:NR_spect} Non-resonant TERS spectra (in arb.u.) for different positions (columns) and field strengths (rows).}
\end{figure*}

Looking at the TERS spectra (Fig.~\ref{fig:NR_spect}) for the three selected positions and possible external electric fields, it is evident that in case of the 'middle' position (Fig.~\ref{fig:NR_spect}b), e), and h)), the field's influence is relatively weak.
The peak with the highest intensity belongs to the same normal mode in all cases, and its intensity stays rougly the same, regardless of the external field. 
However, small variations in the spectra are visible, such as the second-highest peak at \SI{850}{\per\centi\meter} slightly losing intensity in \textbf{SnPc}+E and gaining it in \textbf{SnPc}-E.

For the 'center' and 'edge' positions, the changes are much more pronounced. 
In the 'center', without field, three peaks are prominent in the TERS spectra (Fig.~\ref{fig:NR_spect}d)).
For the \textbf{SnPc-Ag}+E case, those three peaks stay relevant, though their relative intensities change.
Furthermore, the overall intensity of the spectrum increases by more than one order of magnitude.
In contrast, for the reversed field direction, \textbf{SnPc-Ag}-E, most of the spectrum stays at roughly the same intensity as without field, with two of the three prominent peaks mostly vanishing.

At the 'edge' position, similar phenomena are observed.
For the positive field direction, the overall intensity grows, especially for peaks that were already dominating the spectrum, like the one at \SI{1540}{\per\centi\meter}.
The negative field direction also shows a slight increase in the overall intensity in this case, mostly preserving the relative intensities across the spectrum compared to the field-free case.

\subsection{Resonant}

If the energy of the incident light matches that of a dipole-allowed electronic transition, the molecule's response is no longer governed by the (non-resonant) Raman effect, but rather by the resonant one.
In this case, the signal intensity is proportional to the square of the derivative of the transition polarizability, see eq.~\ref{eq:int}.
Therefore, according to eq.~\ref{eq:alpha}, two major contributors to a strong resonant Raman response are the transition dipole moment (in $z$-direction) of the state in resonance and its excitation energy.

\begin{figure*}
 \includegraphics[width=.85\textwidth]{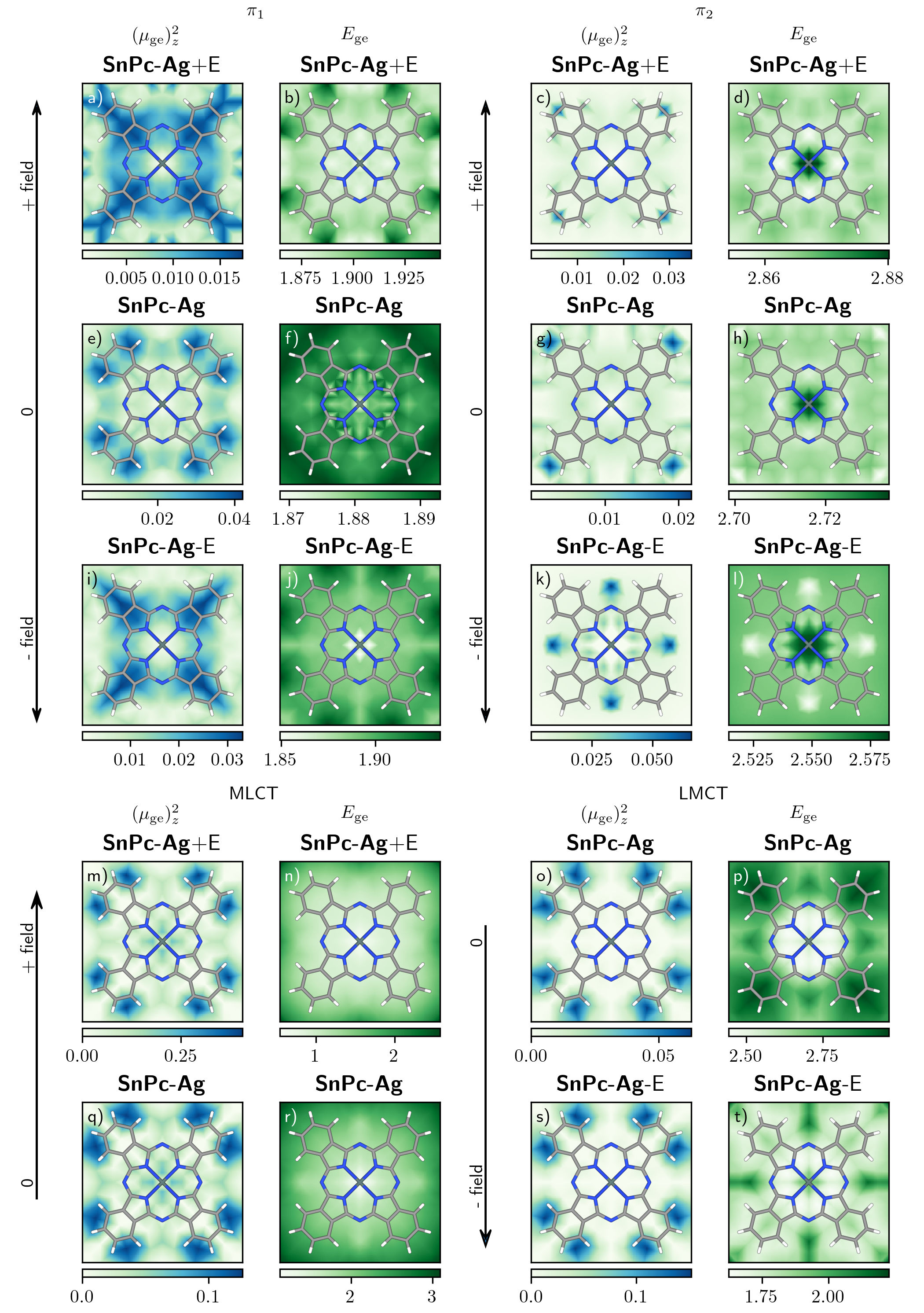}
 \caption{\label{fig:E-mu} Excitation energies (in \si{\electronvolt}) and squared transition dipole moments in $z$-direction (in \si{e\angstrom}). 
 The shown $\pi-\pi^*$ transitions always occur in near-degenerate pairs, for simplicity the displayed energies are averaged between the two, while the transition dipole moments are combined.
 MLCT and LMCT states are not shown for -E and +E, respectively, as these electronic transitions are outside the considered energy window for resonant irradiation.}
\end{figure*}

Fundamentally, there are three different classes of excited states to consider here: 
local excitations of the \acrshort{SnPc} molecule, charge-transfer excitations (in either direction) between the silver tip and the molecule, and local excitations on the silver tip (plasmons for larger clusters of metalic atoms).
The last type of excitation will not be discussed here, as the accurate description of plasmons is not possible with the single-atom model employed for the tip herein. 

In line with previous results\cite{Fiederling_Nanoscale_2020, Fiederling_J.Chem.Phys._2021}, local excitations on the molecule (mostly of $\pi-\pi^*$ character here) are relatively insensitive to tip positioning and applied electric fields, 
both in terms of excitation energy and transition dipole moment, as Fig.~\ref{fig:E-mu} shows.
Typically, these states are also polarized in the molecular plane, which makes them nearly invisible for detection in $z$-direction.
The \acrshort{SnPc} molecule is not wholly planar, therefore these $\pi-\pi^*$ transitions might contribute to signal strength (and therefore sub-molecular resolution) at the slightly upwards-angled edges of the molecule, 
but are otherwise only of minor importance for signal intensity and resolution.

In case of \acrshort{CE}, charge-transfer states between molecule and tip exist (in the observed energy window) in both directions, \textit{i.e.} as \acrfull{MLCT} and \acrfull{LMCT}.
For \acrshort{FULL}, depending on the field direction, one of them gets shifted to lower energies (on average about \SI{0.9}{\electronvolt}), while the other one is shifted significantly higher, outside the observed energy window, 
therefore Fig.~\ref{fig:E-mu} only shows maps for two CT states, respectively.
In contrast to the $\pi-\pi^*$ transitions, the excitation energies are highly dependent on the tip position, regardless of the applied electric field.
The field therefore influences the resolution through strong shifts in excitation energies, where mostly the regions with matching resonance show strong signals. 
However, the transition dipole moment shows a remarkable independence from tip-position and even somewhat from the direction of the charge transfer, while its magnitude changes with field strength. 
Here, the electric fields only play a role in enhancing the signal strength, and do not affect the resolution.

\begin{figure*}
 \includegraphics[width=\textwidth]{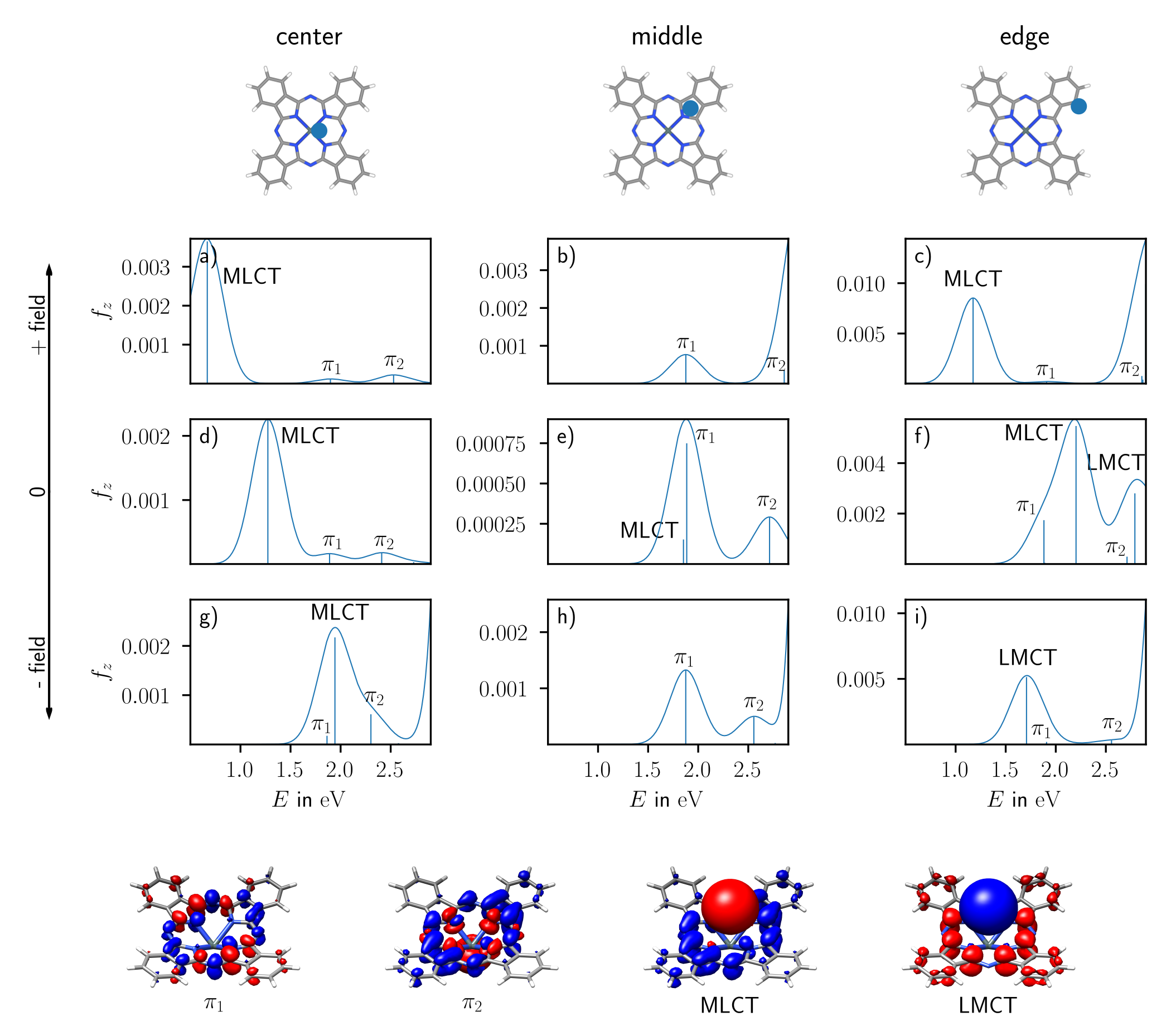}
 \caption{ \label{fig:uvvis} $z$-polarized oscillator strengths, broadened to simulate UV/vis spectra for different positions (columns) and field strengths (rows). 
 Electronic character labeled and indicated exemplarily for the central position in the absence of an electric field.}
 \end{figure*}

The same general behaviour of excited states can be oberved in the UV/vis spectra (Fig.~\ref{fig:uvvis}). 
For the 'middle' position (Fig.~\ref{fig:uvvis}b), e), and h)), mostly only the two sets of near-degenerate $\pi-\pi^*$ states are accessible in the visible region. 
As mentioned above, they are mostly $xy$-polarized, and the transition dipole moment vector has only a small component in $z$-direction to be visible here. 
The molecule-centered $\pi-\pi^*$ states are relatively unaffected by the external fields present in b) and h), both in terms of oscillator strength and excitation energy. 
Charge-transfer states do exist, but are dark ($f_z\approx 0$) at this position.

At the 'center' and 'edge' positions, the $\pi-\pi^*$ states behave accordingly, but are overshadowed by charge-transfer states, that therefore govern the resonant regime.
Already without any external fields, excitation energy and oscillator strength of these states are highly dependent on the tip's position, as we could previously\cite{Fiederling_Nanoscale_2020} show. 
With fields, the CT states' oscillator strengths (and squared transition dipole moments, see Fig.~\ref{fig:E-mu}) are increased slightly in case of \textbf{SnPc-Ag}-E  to moderately (\textbf{SnPc-Ag}+E). 
Additionally, the excitation energies are shifted according to field and charge transfer direction. 
The highest peaks in the field-free case (see d) and f)) are states with electron transfer from the Ag tip to the molecule and are therefore stabilized in \textbf{SnPc-Ag}+E (negative charge above Ag), 
as the respective peaks are shifted to lower excitations energies in a) and c). 
Similarly, in case of \textbf{SnPc-Ag}-E, this charge-transfer state is shifted to higher excitation energy in g) and replaced by another state with opposite charge transfer character (\textit{i.e.} from molecule to tip) in i). 
Due to the enormous sensitivity of the resonance Raman signal on the tip position, the applied field (direction), and excitation energy, we decided not to compare the obtained resonant TERS maps in detail.

\section{Conclusion}

This contribution extends our previous model of a possible TERS experiment further by incorporating inhomogeneous (static) electric fields from a plasmonic nanoparticle, advancing our understanding beyond the chemical effect. 
The electric fields, introduced through sets of point charges, represent examplarily the extreme values reached during the oscillation of the plasmon. 
Our sample molecule, \acrfull{SnPc}, is scanned by a single silver atom and accompanying point charges, mimicking the chemical and electromagnetic interaction with the plasmonic tip in a full quantum mechanical setup. 
The system's response is compared to the same setup without either the silver atom, only showing the effect of the fields, and the \textbf{SnPc-Ag} system, without any field contributions.
This allows us to understand both effects independently and their interaction, in particular regarding signal intensity and possible sub-molecular resolution.
We described the system using \acrshort{DFT}, \acrshort{TDDFT}, and FEM, while registering the Raman response only in $z$-direction; though our approach is easily generalizable to different levels of theory and other illumination-observation geometries.

In case of non-resonant conditions, 
we observed a slight increase in signal intensity by introducing the electric fields to the \acrshort{SnPc} molecule, with relatively uniform change over different tip positions, only slightly highlighting either the center or the edge of the map.
However, the inclusion of the silver atom at the tip's apex introduces strong differences over the whole Raman map, hinting at a sub-molecular resolution.
The combined systems exhibit the same features from the one only including the silver tip, folded together with the signal increase from the field, along with the highlighting of different features (center / edge).
We rationalize this through changing partial charges at the tip atom, reacting to the different chemical environments over different parts of the molecule.

Under resonant conditions, we identify charge-transfer states between tip and molecule as the major contributors to high lateral resolution and signal intensity. 
The excitation energy of these states is highly dependent on tip position as well as field direction and strength. 
However, the transition dipole moment, an important factor for the resonance Raman intensity, is only dependent on the tip position, and therefore the chemical environment. 
Incorporating electric fields only increases the overall magnitude of the transition dipole moment, not its pattern.

Both non-resonant and resonant cases hint at the underlying idea that the chemical interaction between molecule and tip atom shapes the resolution, while the strong electric fields mainly increase signal intensity and function as a means to highlight certain chemical features, \textit{e.g.} through choice of excitation wavelength.
Nevertheless, the chemical contribution and the field effect are strongly influencing each other and should not be considered seperately.

Several additional factors could not be included in this model so far, including any chemical interaction with the substrate, as well as dynamic electric fields from the plasmon, and any sort of 'backwards' interaction from molecule to plasmon. 

\section*{acknowledgements}

 KF thanks Karl Michael Ziems for fruitful discussions. 
The authors gratefully acknowledge funding from the European Research Council (ERC) under the European's Horizon 2020 research and innovation programme 
  -- QUEM-CHEM (grant number 772676), ``Time- and space-resolved ultrafast dynamics in molecular plasmonic hybrid systems``.
  This work was further funded by the Deutsche Forschungsgemeinschaft (DFG, German Research Foundation) -- CRC 1375 NOA (project number 398816777), project A4. 
  All calculations were performed at the Uni\-ver\-si\-täts\-re\-chen\-zen\-trum of the Friedrich Schiller University Jena.

\bibliography{ters-fields}

\begin{thebibliography}{66}%
\makeatletter
\providecommand \@ifxundefined [1]{%
 \@ifx{#1\undefined}
}%
\providecommand \@ifnum [1]{%
 \ifnum #1\expandafter \@firstoftwo
 \else \expandafter \@secondoftwo
 \fi
}%
\providecommand \@ifx [1]{%
 \ifx #1\expandafter \@firstoftwo
 \else \expandafter \@secondoftwo
 \fi
}%
\providecommand \natexlab [1]{#1}%
\providecommand \enquote  [1]{``#1''}%
\providecommand \bibnamefont  [1]{#1}%
\providecommand \bibfnamefont [1]{#1}%
\providecommand \citenamefont [1]{#1}%
\providecommand \href@noop [0]{\@secondoftwo}%
\providecommand \href [0]{\begingroup \@sanitize@url \@href}%
\providecommand \@href[1]{\@@startlink{#1}\@@href}%
\providecommand \@@href[1]{\endgroup#1\@@endlink}%
\providecommand \@sanitize@url [0]{\catcode `\\12\catcode `\$12\catcode
  `\&12\catcode `\#12\catcode `\^12\catcode `\_12\catcode `\%12\relax}%
\providecommand \@@startlink[1]{}%
\providecommand \@@endlink[0]{}%
\providecommand \url  [0]{\begingroup\@sanitize@url \@url }%
\providecommand \@url [1]{\endgroup\@href {#1}{\urlprefix }}%
\providecommand \urlprefix  [0]{URL }%
\providecommand \Eprint [0]{\href }%
\providecommand \doibase [0]{https://doi.org/}%
\providecommand \selectlanguage [0]{\@gobble}%
\providecommand \bibinfo  [0]{\@secondoftwo}%
\providecommand \bibfield  [0]{\@secondoftwo}%
\providecommand \translation [1]{[#1]}%
\providecommand \BibitemOpen [0]{}%
\providecommand \bibitemStop [0]{}%
\providecommand \bibitemNoStop [0]{.\EOS\space}%
\providecommand \EOS [0]{\spacefactor3000\relax}%
\providecommand \BibitemShut  [1]{\csname bibitem#1\endcsname}%
\let\auto@bib@innerbib\@empty
\bibitem [{\citenamefont {St{\"o}ckle}\ \emph {et~al.}(2000)\citenamefont
  {St{\"o}ckle}, \citenamefont {Suh}, \citenamefont {Deckert},\ and\
  \citenamefont {Zenobi}}]{Stockle_ChemicalPhysicsLetters_2000}%
  \BibitemOpen
  \bibfield  {author} {\bibinfo {author} {\bibfnamefont {R.~M.}\ \bibnamefont
  {St{\"o}ckle}}, \bibinfo {author} {\bibfnamefont {Y.~D.}\ \bibnamefont
  {Suh}}, \bibinfo {author} {\bibfnamefont {V.}~\bibnamefont {Deckert}},\ and\
  \bibinfo {author} {\bibfnamefont {R.}~\bibnamefont {Zenobi}},\ }\bibfield
  {title} {\enquote {\bibinfo {title} {Nanoscale chemical analysis by
  tip-enhanced {{Raman}} spectroscopy},}\ }\href
  {https://doi.org/10.1016/S0009-2614(99)01451-7} {\bibfield  {journal}
  {\bibinfo  {journal} {Chemical Physics Letters}\ }\textbf {\bibinfo {volume}
  {318}},\ \bibinfo {pages} {131--136} (\bibinfo {year} {2000})}\BibitemShut
  {NoStop}%
\bibitem [{\citenamefont {Hayazawa}\ \emph {et~al.}(2000)\citenamefont
  {Hayazawa}, \citenamefont {Inouye}, \citenamefont {Sekkat},\ and\
  \citenamefont {Kawata}}]{Hayazawa_OpticsCommunications_2000}%
  \BibitemOpen
  \bibfield  {author} {\bibinfo {author} {\bibfnamefont {N.}~\bibnamefont
  {Hayazawa}}, \bibinfo {author} {\bibfnamefont {Y.}~\bibnamefont {Inouye}},
  \bibinfo {author} {\bibfnamefont {Z.}~\bibnamefont {Sekkat}},\ and\ \bibinfo
  {author} {\bibfnamefont {S.}~\bibnamefont {Kawata}},\ }\bibfield  {title}
  {\enquote {\bibinfo {title} {Metallized tip amplification of near-field
  {{Raman}} scattering},}\ }\href
  {https://doi.org/10.1016/S0030-4018(00)00894-4} {\bibfield  {journal}
  {\bibinfo  {journal} {Optics Communications}\ }\textbf {\bibinfo {volume}
  {183}},\ \bibinfo {pages} {333--336} (\bibinfo {year} {2000})}\BibitemShut
  {NoStop}%
\bibitem [{\citenamefont {Anderson}(2000)}]{Anderson_Appl.Phys.Lett._2000}%
  \BibitemOpen
  \bibfield  {author} {\bibinfo {author} {\bibfnamefont {M.~S.}\ \bibnamefont
  {Anderson}},\ }\bibfield  {title} {\enquote {\bibinfo {title} {Locally
  enhanced {{Raman}} spectroscopy with an atomic force microscope},}\ }\href
  {https://doi.org/10.1063/1.126546} {\bibfield  {journal} {\bibinfo  {journal}
  {Applied Physics Letters}\ }\textbf {\bibinfo {volume} {76}},\ \bibinfo
  {pages} {3130--3132} (\bibinfo {year} {2000})}\BibitemShut {NoStop}%
\bibitem [{\citenamefont {Zhang}\ \emph {et~al.}(2013)\citenamefont {Zhang},
  \citenamefont {Zhang}, \citenamefont {Dong}, \citenamefont {Jiang},
  \citenamefont {Zhang}, \citenamefont {Chen}, \citenamefont {Zhang},
  \citenamefont {Liao}, \citenamefont {Aizpurua}, \citenamefont {Luo},
  \citenamefont {Yang},\ and\ \citenamefont {Hou}}]{Zhang_Nature_2013}%
  \BibitemOpen
  \bibfield  {author} {\bibinfo {author} {\bibfnamefont {R.}~\bibnamefont
  {Zhang}}, \bibinfo {author} {\bibfnamefont {Y.}~\bibnamefont {Zhang}},
  \bibinfo {author} {\bibfnamefont {Z.~C.}\ \bibnamefont {Dong}}, \bibinfo
  {author} {\bibfnamefont {S.}~\bibnamefont {Jiang}}, \bibinfo {author}
  {\bibfnamefont {C.}~\bibnamefont {Zhang}}, \bibinfo {author} {\bibfnamefont
  {L.~G.}\ \bibnamefont {Chen}}, \bibinfo {author} {\bibfnamefont
  {L.}~\bibnamefont {Zhang}}, \bibinfo {author} {\bibfnamefont
  {Y.}~\bibnamefont {Liao}}, \bibinfo {author} {\bibfnamefont {J.}~\bibnamefont
  {Aizpurua}}, \bibinfo {author} {\bibfnamefont {Y.}~\bibnamefont {Luo}},
  \bibinfo {author} {\bibfnamefont {J.~L.}\ \bibnamefont {Yang}},\ and\
  \bibinfo {author} {\bibfnamefont {J.~G.}\ \bibnamefont {Hou}},\ }\bibfield
  {title} {\enquote {\bibinfo {title} {Chemical mapping of a single molecule by
  plasmon-enhanced {{Raman}} scattering},}\ }\href
  {https://doi.org/10.1038/nature12151} {\bibfield  {journal} {\bibinfo
  {journal} {Nature}\ }\textbf {\bibinfo {volume} {498}},\ \bibinfo {pages}
  {82--86} (\bibinfo {year} {2013})}\BibitemShut {NoStop}%
\bibitem [{\citenamefont {Fang}\ \emph {et~al.}(2014)\citenamefont {Fang},
  \citenamefont {Zhang}, \citenamefont {Chen},\ and\ \citenamefont
  {Sun}}]{Fang_Phys.Chem.Chem.Phys._2014}%
  \BibitemOpen
  \bibfield  {author} {\bibinfo {author} {\bibfnamefont {Y.}~\bibnamefont
  {Fang}}, \bibinfo {author} {\bibfnamefont {Z.}~\bibnamefont {Zhang}},
  \bibinfo {author} {\bibfnamefont {L.}~\bibnamefont {Chen}},\ and\ \bibinfo
  {author} {\bibfnamefont {M.}~\bibnamefont {Sun}},\ }\bibfield  {title}
  {\enquote {\bibinfo {title} {Near field plasmonic gradient effects on high
  vacuum tip-enhanced {{Raman}} spectroscopy},}\ }\href
  {https://doi.org/10.1039/C4CP03871A} {\bibfield  {journal} {\bibinfo
  {journal} {Physical Chemistry Chemical Physics}\ }\textbf {\bibinfo {volume}
  {17}},\ \bibinfo {pages} {783--794} (\bibinfo {year} {2014})}\BibitemShut
  {NoStop}%
\bibitem [{\citenamefont {Klingsporn}\ \emph {et~al.}(2014)\citenamefont
  {Klingsporn}, \citenamefont {Jiang}, \citenamefont {Pozzi}, \citenamefont
  {Sonntag}, \citenamefont {Chulhai}, \citenamefont {Seideman}, \citenamefont
  {Jensen}, \citenamefont {Hersam},\ and\ \citenamefont
  {Duyne}}]{Klingsporn_J.Am.Chem.Soc._2014}%
  \BibitemOpen
  \bibfield  {author} {\bibinfo {author} {\bibfnamefont {J.~M.}\ \bibnamefont
  {Klingsporn}}, \bibinfo {author} {\bibfnamefont {N.}~\bibnamefont {Jiang}},
  \bibinfo {author} {\bibfnamefont {E.~A.}\ \bibnamefont {Pozzi}}, \bibinfo
  {author} {\bibfnamefont {M.~D.}\ \bibnamefont {Sonntag}}, \bibinfo {author}
  {\bibfnamefont {D.}~\bibnamefont {Chulhai}}, \bibinfo {author} {\bibfnamefont
  {T.}~\bibnamefont {Seideman}}, \bibinfo {author} {\bibfnamefont
  {L.}~\bibnamefont {Jensen}}, \bibinfo {author} {\bibfnamefont {M.~C.}\
  \bibnamefont {Hersam}},\ and\ \bibinfo {author} {\bibfnamefont {R.~P.~V.}\
  \bibnamefont {Duyne}},\ }\bibfield  {title} {\enquote {\bibinfo {title}
  {Intramolecular {{Insight}} into {{Adsorbate}}\textendash{{Substrate
  Interactions}} via {{Low-Temperature}}, {{Ultrahigh-Vacuum Tip-Enhanced Raman
  Spectroscopy}}},}\ }\href {https://doi.org/10.1021/ja411899k} {\bibfield
  {journal} {\bibinfo  {journal} {Journal of the American Chemical Society}\
  }\textbf {\bibinfo {volume} {136}},\ \bibinfo {pages} {3881--3887} (\bibinfo
  {year} {2014})}\BibitemShut {NoStop}%
\bibitem [{\citenamefont {Sun}\ \emph {et~al.}(2014)\citenamefont {Sun},
  \citenamefont {Zhang}, \citenamefont {Chen}, \citenamefont {Sheng},\ and\
  \citenamefont {Xu}}]{Sun_Adv.Opt.Mater._2014}%
  \BibitemOpen
  \bibfield  {author} {\bibinfo {author} {\bibfnamefont {M.}~\bibnamefont
  {Sun}}, \bibinfo {author} {\bibfnamefont {Z.}~\bibnamefont {Zhang}}, \bibinfo
  {author} {\bibfnamefont {L.}~\bibnamefont {Chen}}, \bibinfo {author}
  {\bibfnamefont {S.}~\bibnamefont {Sheng}},\ and\ \bibinfo {author}
  {\bibfnamefont {H.}~\bibnamefont {Xu}},\ }\bibfield  {title} {\enquote
  {\bibinfo {title} {Plasmonic {{Gradient Effects}} on {{High Vacuum
  Tip-Enhanced Raman Spectroscopy}}},}\ }\href
  {https://doi.org/10.1002/adom.201300296} {\bibfield  {journal} {\bibinfo
  {journal} {Advanced Optical Materials}\ }\textbf {\bibinfo {volume} {2}},\
  \bibinfo {pages} {74--80} (\bibinfo {year} {2014})}\BibitemShut {NoStop}%
\bibitem [{\citenamefont {He}\ \emph {et~al.}(2019)\citenamefont {He},
  \citenamefont {Han}, \citenamefont {Kizer}, \citenamefont {Linhardt},
  \citenamefont {Wang}, \citenamefont {Sinyukov}, \citenamefont {Wang},
  \citenamefont {Deckert}, \citenamefont {Sokolov}, \citenamefont {Hu},\ and\
  \citenamefont {Scully}}]{He_J.Am.Chem.Soc._2019}%
  \BibitemOpen
  \bibfield  {author} {\bibinfo {author} {\bibfnamefont {Z.}~\bibnamefont
  {He}}, \bibinfo {author} {\bibfnamefont {Z.}~\bibnamefont {Han}}, \bibinfo
  {author} {\bibfnamefont {M.}~\bibnamefont {Kizer}}, \bibinfo {author}
  {\bibfnamefont {R.~J.}\ \bibnamefont {Linhardt}}, \bibinfo {author}
  {\bibfnamefont {X.}~\bibnamefont {Wang}}, \bibinfo {author} {\bibfnamefont
  {A.~M.}\ \bibnamefont {Sinyukov}}, \bibinfo {author} {\bibfnamefont
  {J.}~\bibnamefont {Wang}}, \bibinfo {author} {\bibfnamefont {V.}~\bibnamefont
  {Deckert}}, \bibinfo {author} {\bibfnamefont {A.~V.}\ \bibnamefont
  {Sokolov}}, \bibinfo {author} {\bibfnamefont {J.}~\bibnamefont {Hu}},\ and\
  \bibinfo {author} {\bibfnamefont {M.~O.}\ \bibnamefont {Scully}},\ }\bibfield
   {title} {\enquote {\bibinfo {title} {Tip-{{Enhanced Raman Imaging}} of
  {{Single-Stranded DNA}} with {{Single Base Resolution}}},}\ }\href
  {https://doi.org/10.1021/jacs.8b11506} {\bibfield  {journal} {\bibinfo
  {journal} {Journal of the American Chemical Society}\ }\textbf {\bibinfo
  {volume} {141}},\ \bibinfo {pages} {753--757} (\bibinfo {year}
  {2019})}\BibitemShut {NoStop}%
\bibitem [{\citenamefont {Jiang}\ \emph {et~al.}(2015)\citenamefont {Jiang},
  \citenamefont {Zhang}, \citenamefont {Zhang}, \citenamefont {Hu},
  \citenamefont {Liao}, \citenamefont {Luo}, \citenamefont {Yang},
  \citenamefont {Dong},\ and\ \citenamefont
  {Hou}}]{Jiang_Nat.Nanotechnol._2015}%
  \BibitemOpen
  \bibfield  {author} {\bibinfo {author} {\bibfnamefont {S.}~\bibnamefont
  {Jiang}}, \bibinfo {author} {\bibfnamefont {Y.}~\bibnamefont {Zhang}},
  \bibinfo {author} {\bibfnamefont {R.}~\bibnamefont {Zhang}}, \bibinfo
  {author} {\bibfnamefont {C.}~\bibnamefont {Hu}}, \bibinfo {author}
  {\bibfnamefont {M.}~\bibnamefont {Liao}}, \bibinfo {author} {\bibfnamefont
  {Y.}~\bibnamefont {Luo}}, \bibinfo {author} {\bibfnamefont {J.}~\bibnamefont
  {Yang}}, \bibinfo {author} {\bibfnamefont {Z.}~\bibnamefont {Dong}},\ and\
  \bibinfo {author} {\bibfnamefont {J.~G.}\ \bibnamefont {Hou}},\ }\bibfield
  {title} {\enquote {\bibinfo {title} {Distinguishing adjacent molecules on a
  surface using plasmon-enhanced {{Raman}} scattering},}\ }\href
  {https://doi.org/10.1038/nnano.2015.170} {\bibfield  {journal} {\bibinfo
  {journal} {Nature Nanotechnology}\ }\textbf {\bibinfo {volume} {10}},\
  \bibinfo {pages} {865--869} (\bibinfo {year} {2015})}\BibitemShut {NoStop}%
\bibitem [{\citenamefont {Lee}\ \emph {et~al.}(2019)\citenamefont {Lee},
  \citenamefont {Crampton}, \citenamefont {Tallarida},\ and\ \citenamefont
  {Apkarian}}]{Lee_Nature_2019}%
  \BibitemOpen
  \bibfield  {author} {\bibinfo {author} {\bibfnamefont {J.}~\bibnamefont
  {Lee}}, \bibinfo {author} {\bibfnamefont {K.~T.}\ \bibnamefont {Crampton}},
  \bibinfo {author} {\bibfnamefont {N.}~\bibnamefont {Tallarida}},\ and\
  \bibinfo {author} {\bibfnamefont {V.~A.}\ \bibnamefont {Apkarian}},\
  }\bibfield  {title} {\enquote {\bibinfo {title} {Visualizing vibrational
  normal modes of a single molecule with atomically confined light},}\ }\href
  {https://doi.org/10.1038/s41586-019-1059-9} {\bibfield  {journal} {\bibinfo
  {journal} {Nature}\ }\textbf {\bibinfo {volume} {568}},\ \bibinfo {pages}
  {78--82} (\bibinfo {year} {2019})}\BibitemShut {NoStop}%
\bibitem [{\citenamefont {{Richard-Lacroix}}\ \emph {et~al.}(2017)\citenamefont
  {{Richard-Lacroix}}, \citenamefont {Zhang}, \citenamefont {Dong},\ and\
  \citenamefont {Deckert}}]{Richard-Lacroix_Chem.Soc.Rev._2017}%
  \BibitemOpen
  \bibfield  {author} {\bibinfo {author} {\bibfnamefont {M.}~\bibnamefont
  {{Richard-Lacroix}}}, \bibinfo {author} {\bibfnamefont {Y.}~\bibnamefont
  {Zhang}}, \bibinfo {author} {\bibfnamefont {Z.}~\bibnamefont {Dong}},\ and\
  \bibinfo {author} {\bibfnamefont {V.}~\bibnamefont {Deckert}},\ }\bibfield
  {title} {\enquote {\bibinfo {title} {Mastering high resolution tip-enhanced
  {{Raman}} spectroscopy: Towards a shift of perception},}\ }\href
  {https://doi.org/10.1039/C7CS00203C} {\bibfield  {journal} {\bibinfo
  {journal} {Chemical Society Reviews}\ }\textbf {\bibinfo {volume} {46}},\
  \bibinfo {pages} {3922--3944} (\bibinfo {year} {2017})}\BibitemShut {NoStop}%
\bibitem [{\citenamefont {Fleischmann}, \citenamefont {Hendra},\ and\
  \citenamefont {McQuillan}(1974)}]{Fleischmann_ChemicalPhysicsLetters_1974}%
  \BibitemOpen
  \bibfield  {author} {\bibinfo {author} {\bibfnamefont {M.}~\bibnamefont
  {Fleischmann}}, \bibinfo {author} {\bibfnamefont {P.~J.}\ \bibnamefont
  {Hendra}},\ and\ \bibinfo {author} {\bibfnamefont {A.~J.}\ \bibnamefont
  {McQuillan}},\ }\bibfield  {title} {\enquote {\bibinfo {title} {Raman spectra
  of pyridine adsorbed at a silver electrode},}\ }\href
  {https://doi.org/10.1016/0009-2614(74)85388-1} {\bibfield  {journal}
  {\bibinfo  {journal} {Chemical Physics Letters}\ }\textbf {\bibinfo {volume}
  {26}},\ \bibinfo {pages} {163--166} (\bibinfo {year} {1974})}\BibitemShut
  {NoStop}%
\bibitem [{\citenamefont {Albrecht}\ and\ \citenamefont
  {Creighton}(1977)}]{Albrecht_J.Am.Chem.Soc._1977}%
  \BibitemOpen
  \bibfield  {author} {\bibinfo {author} {\bibfnamefont {M.~G.}\ \bibnamefont
  {Albrecht}}\ and\ \bibinfo {author} {\bibfnamefont {J.~A.}\ \bibnamefont
  {Creighton}},\ }\bibfield  {title} {\enquote {\bibinfo {title} {Anomalously
  intense {{Raman}} spectra of pyridine at a silver electrode},}\ }\href
  {https://doi.org/10.1021/ja00457a071} {\bibfield  {journal} {\bibinfo
  {journal} {Journal of the American Chemical Society}\ }\textbf {\bibinfo
  {volume} {99}},\ \bibinfo {pages} {5215--5217} (\bibinfo {year}
  {1977})}\BibitemShut {NoStop}%
\bibitem [{\citenamefont {Jeanmaire}\ and\ \citenamefont
  {Van~Duyne}(1977)}]{Jeanmaire_JournalofElectroanalyticalChemistryandInterfacialElectrochemistry_1977}%
  \BibitemOpen
  \bibfield  {author} {\bibinfo {author} {\bibfnamefont {D.~L.}\ \bibnamefont
  {Jeanmaire}}\ and\ \bibinfo {author} {\bibfnamefont {R.~P.}\ \bibnamefont
  {Van~Duyne}},\ }\bibfield  {title} {\enquote {\bibinfo {title} {Surface raman
  spectroelectrochemistry: {{Part I}}. {{Heterocyclic}}, aromatic, and
  aliphatic amines adsorbed on the anodized silver electrode},}\ }\href
  {https://doi.org/10.1016/S0022-0728(77)80224-6} {\bibfield  {journal}
  {\bibinfo  {journal} {Journal of Electroanalytical Chemistry and Interfacial
  Electrochemistry}\ }\textbf {\bibinfo {volume} {84}},\ \bibinfo {pages}
  {1--20} (\bibinfo {year} {1977})}\BibitemShut {NoStop}%
\bibitem [{\citenamefont {Langer}\ \emph {et~al.}(2020)\citenamefont {Langer},
  \citenamefont {{Jimenez de Aberasturi}}, \citenamefont {Aizpurua},
  \citenamefont {{Alvarez-Puebla}}, \citenamefont {Augui{\'e}}, \citenamefont
  {Baumberg}, \citenamefont {Bazan}, \citenamefont {Bell}, \citenamefont
  {Boisen}, \citenamefont {Brolo}, \citenamefont {Choo}, \citenamefont
  {{Cialla-May}}, \citenamefont {Deckert}, \citenamefont {Fabris},
  \citenamefont {Faulds}, \citenamefont {{Garc{\'i}a de Abajo}}, \citenamefont
  {Goodacre}, \citenamefont {Graham}, \citenamefont {Haes}, \citenamefont
  {Haynes}, \citenamefont {Huck}, \citenamefont {Itoh}, \citenamefont
  {K{\"a}ll}, \citenamefont {Kneipp}, \citenamefont {Kotov}, \citenamefont
  {Kuang}, \citenamefont {Le~Ru}, \citenamefont {Lee}, \citenamefont {Li},
  \citenamefont {Ling}, \citenamefont {Maier}, \citenamefont {Mayerh{\"o}fer},
  \citenamefont {Moskovits}, \citenamefont {Murakoshi}, \citenamefont {Nam},
  \citenamefont {Nie}, \citenamefont {Ozaki}, \citenamefont
  {{Pastoriza-Santos}}, \citenamefont {{Perez-Juste}}, \citenamefont {Popp},
  \citenamefont {Pucci}, \citenamefont {Reich}, \citenamefont {Ren},
  \citenamefont {Schatz}, \citenamefont {Shegai}, \citenamefont
  {Schl{\"u}cker}, \citenamefont {Tay}, \citenamefont {Thomas}, \citenamefont
  {Tian}, \citenamefont {Van~Duyne}, \citenamefont {{Vo-Dinh}}, \citenamefont
  {Wang}, \citenamefont {Willets}, \citenamefont {Xu}, \citenamefont {Xu},
  \citenamefont {Xu}, \citenamefont {Yamamoto}, \citenamefont {Zhao},\ and\
  \citenamefont {{Liz-Marz{\'a}n}}}]{Langer_ACSNano_2020}%
  \BibitemOpen
  \bibfield  {author} {\bibinfo {author} {\bibfnamefont {J.}~\bibnamefont
  {Langer}}, \bibinfo {author} {\bibfnamefont {D.}~\bibnamefont {{Jimenez de
  Aberasturi}}}, \bibinfo {author} {\bibfnamefont {J.}~\bibnamefont
  {Aizpurua}}, \bibinfo {author} {\bibfnamefont {R.~A.}\ \bibnamefont
  {{Alvarez-Puebla}}}, \bibinfo {author} {\bibfnamefont {B.}~\bibnamefont
  {Augui{\'e}}}, \bibinfo {author} {\bibfnamefont {J.~J.}\ \bibnamefont
  {Baumberg}}, \bibinfo {author} {\bibfnamefont {G.~C.}\ \bibnamefont {Bazan}},
  \bibinfo {author} {\bibfnamefont {S.~E.~J.}\ \bibnamefont {Bell}}, \bibinfo
  {author} {\bibfnamefont {A.}~\bibnamefont {Boisen}}, \bibinfo {author}
  {\bibfnamefont {A.~G.}\ \bibnamefont {Brolo}}, \bibinfo {author}
  {\bibfnamefont {J.}~\bibnamefont {Choo}}, \bibinfo {author} {\bibfnamefont
  {D.}~\bibnamefont {{Cialla-May}}}, \bibinfo {author} {\bibfnamefont
  {V.}~\bibnamefont {Deckert}}, \bibinfo {author} {\bibfnamefont
  {L.}~\bibnamefont {Fabris}}, \bibinfo {author} {\bibfnamefont
  {K.}~\bibnamefont {Faulds}}, \bibinfo {author} {\bibfnamefont {F.~J.}\
  \bibnamefont {{Garc{\'i}a de Abajo}}}, \bibinfo {author} {\bibfnamefont
  {R.}~\bibnamefont {Goodacre}}, \bibinfo {author} {\bibfnamefont
  {D.}~\bibnamefont {Graham}}, \bibinfo {author} {\bibfnamefont {A.~J.}\
  \bibnamefont {Haes}}, \bibinfo {author} {\bibfnamefont {C.~L.}\ \bibnamefont
  {Haynes}}, \bibinfo {author} {\bibfnamefont {C.}~\bibnamefont {Huck}},
  \bibinfo {author} {\bibfnamefont {T.}~\bibnamefont {Itoh}}, \bibinfo {author}
  {\bibfnamefont {M.}~\bibnamefont {K{\"a}ll}}, \bibinfo {author}
  {\bibfnamefont {J.}~\bibnamefont {Kneipp}}, \bibinfo {author} {\bibfnamefont
  {N.~A.}\ \bibnamefont {Kotov}}, \bibinfo {author} {\bibfnamefont
  {H.}~\bibnamefont {Kuang}}, \bibinfo {author} {\bibfnamefont {E.~C.}\
  \bibnamefont {Le~Ru}}, \bibinfo {author} {\bibfnamefont {H.~K.}\ \bibnamefont
  {Lee}}, \bibinfo {author} {\bibfnamefont {J.-F.}\ \bibnamefont {Li}},
  \bibinfo {author} {\bibfnamefont {X.~Y.}\ \bibnamefont {Ling}}, \bibinfo
  {author} {\bibfnamefont {S.~A.}\ \bibnamefont {Maier}}, \bibinfo {author}
  {\bibfnamefont {T.}~\bibnamefont {Mayerh{\"o}fer}}, \bibinfo {author}
  {\bibfnamefont {M.}~\bibnamefont {Moskovits}}, \bibinfo {author}
  {\bibfnamefont {K.}~\bibnamefont {Murakoshi}}, \bibinfo {author}
  {\bibfnamefont {J.-M.}\ \bibnamefont {Nam}}, \bibinfo {author} {\bibfnamefont
  {S.}~\bibnamefont {Nie}}, \bibinfo {author} {\bibfnamefont {Y.}~\bibnamefont
  {Ozaki}}, \bibinfo {author} {\bibfnamefont {I.}~\bibnamefont
  {{Pastoriza-Santos}}}, \bibinfo {author} {\bibfnamefont {J.}~\bibnamefont
  {{Perez-Juste}}}, \bibinfo {author} {\bibfnamefont {J.}~\bibnamefont {Popp}},
  \bibinfo {author} {\bibfnamefont {A.}~\bibnamefont {Pucci}}, \bibinfo
  {author} {\bibfnamefont {S.}~\bibnamefont {Reich}}, \bibinfo {author}
  {\bibfnamefont {B.}~\bibnamefont {Ren}}, \bibinfo {author} {\bibfnamefont
  {G.~C.}\ \bibnamefont {Schatz}}, \bibinfo {author} {\bibfnamefont
  {T.}~\bibnamefont {Shegai}}, \bibinfo {author} {\bibfnamefont
  {S.}~\bibnamefont {Schl{\"u}cker}}, \bibinfo {author} {\bibfnamefont {L.-L.}\
  \bibnamefont {Tay}}, \bibinfo {author} {\bibfnamefont {K.~G.}\ \bibnamefont
  {Thomas}}, \bibinfo {author} {\bibfnamefont {Z.-Q.}\ \bibnamefont {Tian}},
  \bibinfo {author} {\bibfnamefont {R.~P.}\ \bibnamefont {Van~Duyne}}, \bibinfo
  {author} {\bibfnamefont {T.}~\bibnamefont {{Vo-Dinh}}}, \bibinfo {author}
  {\bibfnamefont {Y.}~\bibnamefont {Wang}}, \bibinfo {author} {\bibfnamefont
  {K.~A.}\ \bibnamefont {Willets}}, \bibinfo {author} {\bibfnamefont
  {C.}~\bibnamefont {Xu}}, \bibinfo {author} {\bibfnamefont {H.}~\bibnamefont
  {Xu}}, \bibinfo {author} {\bibfnamefont {Y.}~\bibnamefont {Xu}}, \bibinfo
  {author} {\bibfnamefont {Y.~S.}\ \bibnamefont {Yamamoto}}, \bibinfo {author}
  {\bibfnamefont {B.}~\bibnamefont {Zhao}},\ and\ \bibinfo {author}
  {\bibfnamefont {L.~M.}\ \bibnamefont {{Liz-Marz{\'a}n}}},\ }\bibfield
  {title} {\enquote {\bibinfo {title} {Present and {{Future}} of
  {{Surface-Enhanced Raman Scattering}}},}\ }\href
  {https://doi.org/10.1021/acsnano.9b04224} {\bibfield  {journal} {\bibinfo
  {journal} {ACS Nano}\ }\textbf {\bibinfo {volume} {14}},\ \bibinfo {pages}
  {28--117} (\bibinfo {year} {2020})}\BibitemShut {NoStop}%
\bibitem [{\citenamefont {Jiang}\ \emph {et~al.}(2012)\citenamefont {Jiang},
  \citenamefont {Foley}, \citenamefont {Klingsporn}, \citenamefont {Sonntag},
  \citenamefont {Valley}, \citenamefont {Dieringer}, \citenamefont {Seideman},
  \citenamefont {Schatz}, \citenamefont {Hersam},\ and\ \citenamefont
  {Van~Duyne}}]{Jiang_NanoLett._2012}%
  \BibitemOpen
  \bibfield  {author} {\bibinfo {author} {\bibfnamefont {N.}~\bibnamefont
  {Jiang}}, \bibinfo {author} {\bibfnamefont {E.~T.}\ \bibnamefont {Foley}},
  \bibinfo {author} {\bibfnamefont {J.~M.}\ \bibnamefont {Klingsporn}},
  \bibinfo {author} {\bibfnamefont {M.~D.}\ \bibnamefont {Sonntag}}, \bibinfo
  {author} {\bibfnamefont {N.~A.}\ \bibnamefont {Valley}}, \bibinfo {author}
  {\bibfnamefont {J.~A.}\ \bibnamefont {Dieringer}}, \bibinfo {author}
  {\bibfnamefont {T.}~\bibnamefont {Seideman}}, \bibinfo {author}
  {\bibfnamefont {G.~C.}\ \bibnamefont {Schatz}}, \bibinfo {author}
  {\bibfnamefont {M.~C.}\ \bibnamefont {Hersam}},\ and\ \bibinfo {author}
  {\bibfnamefont {R.~P.}\ \bibnamefont {Van~Duyne}},\ }\bibfield  {title}
  {\enquote {\bibinfo {title} {Observation of {{Multiple Vibrational Modes}} in
  {{Ultrahigh Vacuum Tip-Enhanced Raman Spectroscopy Combined}} with
  {{Molecular-Resolution Scanning Tunneling Microscopy}}},}\ }\href
  {https://doi.org/10.1021/nl2039925} {\bibfield  {journal} {\bibinfo
  {journal} {Nano Letters}\ }\textbf {\bibinfo {volume} {12}},\ \bibinfo
  {pages} {5061--5067} (\bibinfo {year} {2012})}\BibitemShut {NoStop}%
\bibitem [{\citenamefont {Pettinger}\ \emph {et~al.}(2004)\citenamefont
  {Pettinger}, \citenamefont {Ren}, \citenamefont {Picardi}, \citenamefont
  {Schuster},\ and\ \citenamefont {Ertl}}]{Pettinger_Phys.Rev.Lett._2004}%
  \BibitemOpen
  \bibfield  {author} {\bibinfo {author} {\bibfnamefont {B.}~\bibnamefont
  {Pettinger}}, \bibinfo {author} {\bibfnamefont {B.}~\bibnamefont {Ren}},
  \bibinfo {author} {\bibfnamefont {G.}~\bibnamefont {Picardi}}, \bibinfo
  {author} {\bibfnamefont {R.}~\bibnamefont {Schuster}},\ and\ \bibinfo
  {author} {\bibfnamefont {G.}~\bibnamefont {Ertl}},\ }\bibfield  {title}
  {\enquote {\bibinfo {title} {Nanoscale {{Probing}} of {{Adsorbed Species}} by
  {{Tip-Enhanced Raman Spectroscopy}}},}\ }\href
  {https://doi.org/10.1103/PhysRevLett.92.096101} {\bibfield  {journal}
  {\bibinfo  {journal} {Physical Review Letters}\ }\textbf {\bibinfo {volume}
  {92}},\ \bibinfo {pages} {096101} (\bibinfo {year} {2004})}\BibitemShut
  {NoStop}%
\bibitem [{\citenamefont {Corni}\ and\ \citenamefont
  {Tomasi}(2001{\natexlab{a}})}]{Corni_ChemicalPhysicsLetters_2001}%
  \BibitemOpen
  \bibfield  {author} {\bibinfo {author} {\bibfnamefont {S.}~\bibnamefont
  {Corni}}\ and\ \bibinfo {author} {\bibfnamefont {J.}~\bibnamefont {Tomasi}},\
  }\bibfield  {title} {\enquote {\bibinfo {title} {Theoretical evaluation of
  {{Raman}} spectra and enhancement factors for a molecule adsorbed on a
  complex-shaped metal particle},}\ }\href
  {https://doi.org/10.1016/S0009-2614(01)00582-6} {\bibfield  {journal}
  {\bibinfo  {journal} {Chemical Physics Letters}\ }\textbf {\bibinfo {volume}
  {342}},\ \bibinfo {pages} {135--140} (\bibinfo {year}
  {2001}{\natexlab{a}})}\BibitemShut {NoStop}%
\bibitem [{\citenamefont {Corni}\ and\ \citenamefont
  {Tomasi}(2001{\natexlab{b}})}]{Corni_J.Chem.Phys._2001a}%
  \BibitemOpen
  \bibfield  {author} {\bibinfo {author} {\bibfnamefont {S.}~\bibnamefont
  {Corni}}\ and\ \bibinfo {author} {\bibfnamefont {J.}~\bibnamefont {Tomasi}},\
  }\bibfield  {title} {\enquote {\bibinfo {title} {Surface enhanced {{Raman}}
  scattering from a single molecule adsorbed on a metal particle aggregate:
  {{A}} theoretical study},}\ }\href {https://doi.org/10.1063/1.1428349}
  {\bibfield  {journal} {\bibinfo  {journal} {The Journal of Chemical Physics}\
  }\textbf {\bibinfo {volume} {116}},\ \bibinfo {pages} {1156--1164} (\bibinfo
  {year} {2001}{\natexlab{b}})}\BibitemShut {NoStop}%
\bibitem [{\citenamefont {Chulhai}\ and\ \citenamefont
  {Jensen}(2013)}]{Chulhai_J.Phys.Chem.C_2013}%
  \BibitemOpen
  \bibfield  {author} {\bibinfo {author} {\bibfnamefont {D.~V.}\ \bibnamefont
  {Chulhai}}\ and\ \bibinfo {author} {\bibfnamefont {L.}~\bibnamefont
  {Jensen}},\ }\bibfield  {title} {\enquote {\bibinfo {title} {Determining
  {{Molecular Orientation With Surface-Enhanced Raman Scattering Using
  Inhomogenous Electric Fields}}},}\ }\href {https://doi.org/10.1021/jp4062626}
  {\bibfield  {journal} {\bibinfo  {journal} {The Journal of Physical Chemistry
  C}\ }\textbf {\bibinfo {volume} {117}},\ \bibinfo {pages} {19622--19631}
  (\bibinfo {year} {2013})}\BibitemShut {NoStop}%
\bibitem [{\citenamefont {Chiang}\ \emph {et~al.}(2015)\citenamefont {Chiang},
  \citenamefont {Jiang}, \citenamefont {Chulhai}, \citenamefont {Pozzi},
  \citenamefont {Hersam}, \citenamefont {Jensen}, \citenamefont {Seideman},\
  and\ \citenamefont {Van~Duyne}}]{Chiang_NanoLett._2015}%
  \BibitemOpen
  \bibfield  {author} {\bibinfo {author} {\bibfnamefont {N.}~\bibnamefont
  {Chiang}}, \bibinfo {author} {\bibfnamefont {N.}~\bibnamefont {Jiang}},
  \bibinfo {author} {\bibfnamefont {D.~V.}\ \bibnamefont {Chulhai}}, \bibinfo
  {author} {\bibfnamefont {E.~A.}\ \bibnamefont {Pozzi}}, \bibinfo {author}
  {\bibfnamefont {M.~C.}\ \bibnamefont {Hersam}}, \bibinfo {author}
  {\bibfnamefont {L.}~\bibnamefont {Jensen}}, \bibinfo {author} {\bibfnamefont
  {T.}~\bibnamefont {Seideman}},\ and\ \bibinfo {author} {\bibfnamefont
  {R.~P.}\ \bibnamefont {Van~Duyne}},\ }\bibfield  {title} {\enquote {\bibinfo
  {title} {Molecular-{{Resolution Interrogation}} of a {{Porphyrin Monolayer}}
  by {{Ultrahigh Vacuum Tip-Enhanced Raman}} and {{Fluorescence
  Spectroscopy}}},}\ }\href {https://doi.org/10.1021/acs.nanolett.5b01225}
  {\bibfield  {journal} {\bibinfo  {journal} {Nano Letters}\ }\textbf {\bibinfo
  {volume} {15}},\ \bibinfo {pages} {4114--4120} (\bibinfo {year}
  {2015})}\BibitemShut {NoStop}%
\bibitem [{\citenamefont {Zhang}, \citenamefont {Chen},\ and\ \citenamefont
  {Li}(2015)}]{Zhang_J.Phys.Chem.C_2015}%
  \BibitemOpen
  \bibfield  {author} {\bibinfo {author} {\bibfnamefont {C.}~\bibnamefont
  {Zhang}}, \bibinfo {author} {\bibfnamefont {B.-Q.}\ \bibnamefont {Chen}},\
  and\ \bibinfo {author} {\bibfnamefont {Z.-Y.}\ \bibnamefont {Li}},\
  }\bibfield  {title} {\enquote {\bibinfo {title} {Optical {{Origin}} of
  {{Subnanometer Resolution}} in {{Tip-Enhanced Raman Mapping}}},}\ }\href
  {https://doi.org/10.1021/acs.jpcc.5b02653} {\bibfield  {journal} {\bibinfo
  {journal} {The Journal of Physical Chemistry C}\ }\textbf {\bibinfo {volume}
  {119}},\ \bibinfo {pages} {11858--11871} (\bibinfo {year}
  {2015})}\BibitemShut {NoStop}%
\bibitem [{\citenamefont {Morton}, \citenamefont {Silverstein},\ and\
  \citenamefont {Jensen}(2011)}]{Morton_Chem.Rev._2011}%
  \BibitemOpen
  \bibfield  {author} {\bibinfo {author} {\bibfnamefont {S.~M.}\ \bibnamefont
  {Morton}}, \bibinfo {author} {\bibfnamefont {D.~W.}\ \bibnamefont
  {Silverstein}},\ and\ \bibinfo {author} {\bibfnamefont {L.}~\bibnamefont
  {Jensen}},\ }\bibfield  {title} {\enquote {\bibinfo {title} {Theoretical
  {{Studies}} of {{Plasmonics}} using {{Electronic Structure Methods}}},}\
  }\href {https://doi.org/10.1021/cr100265f} {\bibfield  {journal} {\bibinfo
  {journal} {Chemical Reviews}\ }\textbf {\bibinfo {volume} {111}},\ \bibinfo
  {pages} {3962--3994} (\bibinfo {year} {2011})}\BibitemShut {NoStop}%
\bibitem [{\citenamefont {Thomas}\ \emph {et~al.}(2013)\citenamefont {Thomas},
  \citenamefont {M{\"u}hlig}, \citenamefont {Deckert-Gaudig}, \citenamefont
  {Rockstuhl}, \citenamefont {Deckert},\ and\ \citenamefont
  {Marquetand}}]{Thomas_J.RamanSpectrosc._2013}%
  \BibitemOpen
  \bibfield  {author} {\bibinfo {author} {\bibfnamefont {M.}~\bibnamefont
  {Thomas}}, \bibinfo {author} {\bibfnamefont {S.}~\bibnamefont {M{\"u}hlig}},
  \bibinfo {author} {\bibfnamefont {T.}~\bibnamefont {Deckert-Gaudig}},
  \bibinfo {author} {\bibfnamefont {C.}~\bibnamefont {Rockstuhl}}, \bibinfo
  {author} {\bibfnamefont {V.}~\bibnamefont {Deckert}},\ and\ \bibinfo {author}
  {\bibfnamefont {P.}~\bibnamefont {Marquetand}},\ }\bibfield  {title}
  {\enquote {\bibinfo {title} {Distinguishing chemical and electromagnetic
  enhancement in surface-enhanced {{Raman}} spectra: {{The}} case of
  para-nitrothiophenol},}\ }\href {https://doi.org/10.1002/jrs.4377} {\bibfield
   {journal} {\bibinfo  {journal} {Journal of Raman Spectroscopy}\ }\textbf
  {\bibinfo {volume} {44}},\ \bibinfo {pages} {1497--1505} (\bibinfo {year}
  {2013})}\BibitemShut {NoStop}%
\bibitem [{\citenamefont {Zhang}\ \emph {et~al.}(2014)\citenamefont {Zhang},
  \citenamefont {Feist}, \citenamefont {Rubio}, \citenamefont
  {{Garc{\'i}a-Gonz{\'a}lez}},\ and\ \citenamefont
  {{Garc{\'i}a-Vidal}}}]{Zhang_Phys.Rev.B_2014}%
  \BibitemOpen
  \bibfield  {author} {\bibinfo {author} {\bibfnamefont {P.}~\bibnamefont
  {Zhang}}, \bibinfo {author} {\bibfnamefont {J.}~\bibnamefont {Feist}},
  \bibinfo {author} {\bibfnamefont {A.}~\bibnamefont {Rubio}}, \bibinfo
  {author} {\bibfnamefont {P.}~\bibnamefont {{Garc{\'i}a-Gonz{\'a}lez}}},\ and\
  \bibinfo {author} {\bibfnamefont {F.~J.}\ \bibnamefont
  {{Garc{\'i}a-Vidal}}},\ }\bibfield  {title} {\enquote {\bibinfo {title} {Ab
  initio nanoplasmonics: {{The}} impact of atomic structure},}\ }\href
  {https://doi.org/10.1103/PhysRevB.90.161407} {\bibfield  {journal} {\bibinfo
  {journal} {Physical Review B}\ }\textbf {\bibinfo {volume} {90}},\ \bibinfo
  {pages} {161407} (\bibinfo {year} {2014})}\BibitemShut {NoStop}%
\bibitem [{\citenamefont {Benz}\ \emph {et~al.}(2016)\citenamefont {Benz},
  \citenamefont {Schmidt}, \citenamefont {Dreismann}, \citenamefont
  {Chikkaraddy}, \citenamefont {Zhang}, \citenamefont {Demetriadou},
  \citenamefont {Carnegie}, \citenamefont {Ohadi}, \citenamefont {de~Nijs},
  \citenamefont {Esteban}, \citenamefont {Aizpurua},\ and\ \citenamefont
  {Baumberg}}]{Benz_Science_2016}%
  \BibitemOpen
  \bibfield  {author} {\bibinfo {author} {\bibfnamefont {F.}~\bibnamefont
  {Benz}}, \bibinfo {author} {\bibfnamefont {M.~K.}\ \bibnamefont {Schmidt}},
  \bibinfo {author} {\bibfnamefont {A.}~\bibnamefont {Dreismann}}, \bibinfo
  {author} {\bibfnamefont {R.}~\bibnamefont {Chikkaraddy}}, \bibinfo {author}
  {\bibfnamefont {Y.}~\bibnamefont {Zhang}}, \bibinfo {author} {\bibfnamefont
  {A.}~\bibnamefont {Demetriadou}}, \bibinfo {author} {\bibfnamefont
  {C.}~\bibnamefont {Carnegie}}, \bibinfo {author} {\bibfnamefont
  {H.}~\bibnamefont {Ohadi}}, \bibinfo {author} {\bibfnamefont
  {B.}~\bibnamefont {de~Nijs}}, \bibinfo {author} {\bibfnamefont
  {R.}~\bibnamefont {Esteban}}, \bibinfo {author} {\bibfnamefont
  {J.}~\bibnamefont {Aizpurua}},\ and\ \bibinfo {author} {\bibfnamefont
  {J.~J.}\ \bibnamefont {Baumberg}},\ }\bibfield  {title} {\enquote {\bibinfo
  {title} {Single-molecule optomechanics in ``picocavities''},}\ }\href
  {https://doi.org/10.1126/science.aah5243} {\bibfield  {journal} {\bibinfo
  {journal} {Science}\ }\textbf {\bibinfo {volume} {354}},\ \bibinfo {pages}
  {726--729} (\bibinfo {year} {2016})}\BibitemShut {NoStop}%
\bibitem [{\citenamefont {Trautmann}\ \emph {et~al.}(2016)\citenamefont
  {Trautmann}, \citenamefont {Aizpurua}, \citenamefont {G{\"o}tz},
  \citenamefont {Undisz}, \citenamefont {Dellith}, \citenamefont
  {Schneidewind}, \citenamefont {Rettenmayr},\ and\ \citenamefont
  {Deckert}}]{Trautmann_Nanoscale_2016}%
  \BibitemOpen
  \bibfield  {author} {\bibinfo {author} {\bibfnamefont {S.}~\bibnamefont
  {Trautmann}}, \bibinfo {author} {\bibfnamefont {J.}~\bibnamefont {Aizpurua}},
  \bibinfo {author} {\bibfnamefont {I.}~\bibnamefont {G{\"o}tz}}, \bibinfo
  {author} {\bibfnamefont {A.}~\bibnamefont {Undisz}}, \bibinfo {author}
  {\bibfnamefont {J.}~\bibnamefont {Dellith}}, \bibinfo {author} {\bibfnamefont
  {H.}~\bibnamefont {Schneidewind}}, \bibinfo {author} {\bibfnamefont
  {M.}~\bibnamefont {Rettenmayr}},\ and\ \bibinfo {author} {\bibfnamefont
  {V.}~\bibnamefont {Deckert}},\ }\bibfield  {title} {\enquote {\bibinfo
  {title} {A classical description of subnanometer resolution by atomic
  features in metallic structures},}\ }\href
  {https://doi.org/10.1039/C6NR07560F} {\bibfield  {journal} {\bibinfo
  {journal} {Nanoscale}\ }\textbf {\bibinfo {volume} {9}},\ \bibinfo {pages}
  {391--401} (\bibinfo {year} {2016})}\BibitemShut {NoStop}%
\bibitem [{\citenamefont {Chen}\ and\ \citenamefont
  {Jensen}(2018)}]{Chen_Nanoscale_2018}%
  \BibitemOpen
  \bibfield  {author} {\bibinfo {author} {\bibfnamefont {X.}~\bibnamefont
  {Chen}}\ and\ \bibinfo {author} {\bibfnamefont {L.}~\bibnamefont {Jensen}},\
  }\bibfield  {title} {\enquote {\bibinfo {title} {Morphology dependent
  near-field response in atomistic plasmonic nanocavities},}\ }\href
  {https://doi.org/10.1039/C8NR03029D} {\bibfield  {journal} {\bibinfo
  {journal} {Nanoscale}\ }\textbf {\bibinfo {volume} {10}},\ \bibinfo {pages}
  {11410--11417} (\bibinfo {year} {2018})}\BibitemShut {NoStop}%
\bibitem [{\citenamefont {Urbieta}\ \emph {et~al.}(2018)\citenamefont
  {Urbieta}, \citenamefont {Barbry}, \citenamefont {Zhang}, \citenamefont
  {Koval}, \citenamefont {{S{\'a}nchez-Portal}}, \citenamefont {Zabala},\ and\
  \citenamefont {Aizpurua}}]{Urbieta_ACSNano_2018}%
  \BibitemOpen
  \bibfield  {author} {\bibinfo {author} {\bibfnamefont {M.}~\bibnamefont
  {Urbieta}}, \bibinfo {author} {\bibfnamefont {M.}~\bibnamefont {Barbry}},
  \bibinfo {author} {\bibfnamefont {Y.}~\bibnamefont {Zhang}}, \bibinfo
  {author} {\bibfnamefont {P.}~\bibnamefont {Koval}}, \bibinfo {author}
  {\bibfnamefont {D.}~\bibnamefont {{S{\'a}nchez-Portal}}}, \bibinfo {author}
  {\bibfnamefont {N.}~\bibnamefont {Zabala}},\ and\ \bibinfo {author}
  {\bibfnamefont {J.}~\bibnamefont {Aizpurua}},\ }\bibfield  {title} {\enquote
  {\bibinfo {title} {Atomic-{{Scale Lightning Rod Effect}} in {{Plasmonic
  Picocavities}}: {{A Classical View}} to a {{Quantum Effect}}},}\ }\href
  {https://doi.org/10.1021/acsnano.7b07401} {\bibfield  {journal} {\bibinfo
  {journal} {ACS Nano}\ }\textbf {\bibinfo {volume} {12}},\ \bibinfo {pages}
  {585--595} (\bibinfo {year} {2018})}\BibitemShut {NoStop}%
\bibitem [{\citenamefont {Baumberg}\ \emph {et~al.}(2019)\citenamefont
  {Baumberg}, \citenamefont {Aizpurua}, \citenamefont {Mikkelsen},\ and\
  \citenamefont {Smith}}]{Baumberg_Nat.Mater._2019}%
  \BibitemOpen
  \bibfield  {author} {\bibinfo {author} {\bibfnamefont {J.~J.}\ \bibnamefont
  {Baumberg}}, \bibinfo {author} {\bibfnamefont {J.}~\bibnamefont {Aizpurua}},
  \bibinfo {author} {\bibfnamefont {M.~H.}\ \bibnamefont {Mikkelsen}},\ and\
  \bibinfo {author} {\bibfnamefont {D.~R.}\ \bibnamefont {Smith}},\ }\bibfield
  {title} {\enquote {\bibinfo {title} {Extreme nanophotonics from ultrathin
  metallic gaps},}\ }\href {https://doi.org/10.1038/s41563-019-0290-y}
  {\bibfield  {journal} {\bibinfo  {journal} {Nature Materials}\ }\textbf
  {\bibinfo {volume} {18}},\ \bibinfo {pages} {668--678} (\bibinfo {year}
  {2019})}\BibitemShut {NoStop}%
\bibitem [{\citenamefont {{Zhao}}, \citenamefont {Jensen},\ and\ \citenamefont
  {Schatz}(2006)}]{Zhao_J.Am.Chem.Soc._2006}%
  \BibitemOpen
  \bibfield  {author} {\bibinfo {author} {\bibnamefont {{Zhao}}}, \bibinfo
  {author} {\bibfnamefont {L.}~\bibnamefont {Jensen}},\ and\ \bibinfo {author}
  {\bibfnamefont {G.~C.}\ \bibnamefont {Schatz}},\ }\bibfield  {title}
  {\enquote {\bibinfo {title} {{{Pyridine}}-{{Ag20 Cluster}}:\, {{A Model
  System}} for {{Studying Surface-Enhanced Raman Scattering}}},}\ }\href
  {https://doi.org/10.1021/ja0556326} {\bibfield  {journal} {\bibinfo
  {journal} {Journal of the American Chemical Society}\ }\textbf {\bibinfo
  {volume} {128}},\ \bibinfo {pages} {2911--2919} (\bibinfo {year}
  {2006})}\BibitemShut {NoStop}%
\bibitem [{\citenamefont {Jensen}, \citenamefont {Zhao},\ and\ \citenamefont
  {Schatz}(2007)}]{Jensen_J.Phys.Chem.C_2007}%
  \BibitemOpen
  \bibfield  {author} {\bibinfo {author} {\bibfnamefont {L.}~\bibnamefont
  {Jensen}}, \bibinfo {author} {\bibfnamefont {L.~L.}\ \bibnamefont {Zhao}},\
  and\ \bibinfo {author} {\bibfnamefont {G.~C.}\ \bibnamefont {Schatz}},\
  }\bibfield  {title} {\enquote {\bibinfo {title} {Size-{{Dependence}} of the
  {{Enhanced Raman Scattering}} of {{Pyridine Adsorbed}} on {{Agn}} (n = 2-8,
  20) {{Clusters}}},}\ }\href {https://doi.org/10.1021/jp067634y} {\bibfield
  {journal} {\bibinfo  {journal} {The Journal of Physical Chemistry C}\
  }\textbf {\bibinfo {volume} {111}},\ \bibinfo {pages} {4756--4764} (\bibinfo
  {year} {2007})}\BibitemShut {NoStop}%
\bibitem [{\citenamefont {Jensen}, \citenamefont {Aikens},\ and\ \citenamefont
  {Schatz}(2008)}]{Jensen_Chem.Soc.Rev._2008}%
  \BibitemOpen
  \bibfield  {author} {\bibinfo {author} {\bibfnamefont {L.}~\bibnamefont
  {Jensen}}, \bibinfo {author} {\bibfnamefont {C.~M.}\ \bibnamefont {Aikens}},\
  and\ \bibinfo {author} {\bibfnamefont {G.~C.}\ \bibnamefont {Schatz}},\
  }\bibfield  {title} {\enquote {\bibinfo {title} {Electronic structure methods
  for studying surface-enhanced {{Raman}} scattering},}\ }\href
  {https://doi.org/10.1039/B706023H} {\bibfield  {journal} {\bibinfo  {journal}
  {Chemical Society Reviews}\ }\textbf {\bibinfo {volume} {37}},\ \bibinfo
  {pages} {1061--1073} (\bibinfo {year} {2008})}\BibitemShut {NoStop}%
\bibitem [{\citenamefont {Liu}\ \emph {et~al.}(2009)\citenamefont {Liu},
  \citenamefont {Zhao}, \citenamefont {Li}, \citenamefont {Chen},\ and\
  \citenamefont
  {Sun}}]{Liu_SpectrochimicaActaPartA:MolecularandBiomolecularSpectroscopy_2009}%
  \BibitemOpen
  \bibfield  {author} {\bibinfo {author} {\bibfnamefont {S.}~\bibnamefont
  {Liu}}, \bibinfo {author} {\bibfnamefont {X.}~\bibnamefont {Zhao}}, \bibinfo
  {author} {\bibfnamefont {Y.}~\bibnamefont {Li}}, \bibinfo {author}
  {\bibfnamefont {M.}~\bibnamefont {Chen}},\ and\ \bibinfo {author}
  {\bibfnamefont {M.}~\bibnamefont {Sun}},\ }\bibfield  {title} {\enquote
  {\bibinfo {title} {{{DFT}} study of adsorption site effect on
  surface-enhanced {{Raman}} scattering of neutral and charged
  pyridine\textendash{{Ag4}} complexes},}\ }\href
  {https://doi.org/10.1016/j.saa.2009.02.036} {\bibfield  {journal} {\bibinfo
  {journal} {Spectrochimica Acta Part A: Molecular and Biomolecular
  Spectroscopy}\ }\textbf {\bibinfo {volume} {73}},\ \bibinfo {pages}
  {382--387} (\bibinfo {year} {2009})}\BibitemShut {NoStop}%
\bibitem [{\citenamefont {Valley}\ \emph {et~al.}(2013)\citenamefont {Valley},
  \citenamefont {Greeneltch}, \citenamefont {Van~Duyne},\ and\ \citenamefont
  {Schatz}}]{Valley_J.Phys.Chem.Lett._2013}%
  \BibitemOpen
  \bibfield  {author} {\bibinfo {author} {\bibfnamefont {N.}~\bibnamefont
  {Valley}}, \bibinfo {author} {\bibfnamefont {N.}~\bibnamefont {Greeneltch}},
  \bibinfo {author} {\bibfnamefont {R.~P.}\ \bibnamefont {Van~Duyne}},\ and\
  \bibinfo {author} {\bibfnamefont {G.~C.}\ \bibnamefont {Schatz}},\ }\bibfield
   {title} {\enquote {\bibinfo {title} {A {{Look}} at the {{Origin}} and
  {{Magnitude}} of the {{Chemical Contribution}} to the {{Enhancement
  Mechanism}} of {{Surface-Enhanced Raman Spectroscopy}} ({{SERS}}): {{Theory}}
  and {{Experiment}}},}\ }\href {https://doi.org/10.1021/jz4012383} {\bibfield
  {journal} {\bibinfo  {journal} {The Journal of Physical Chemistry Letters}\
  }\textbf {\bibinfo {volume} {4}},\ \bibinfo {pages} {2599--2604} (\bibinfo
  {year} {2013})}\BibitemShut {NoStop}%
\bibitem [{\citenamefont {Latorre}, \citenamefont {Guthmuller},\ and\
  \citenamefont {Marquetand}(2015)}]{Latorre_Phys.Chem.Chem.Phys._2015}%
  \BibitemOpen
  \bibfield  {author} {\bibinfo {author} {\bibfnamefont {F.}~\bibnamefont
  {Latorre}}, \bibinfo {author} {\bibfnamefont {J.}~\bibnamefont
  {Guthmuller}},\ and\ \bibinfo {author} {\bibfnamefont {P.}~\bibnamefont
  {Marquetand}},\ }\bibfield  {title} {\enquote {\bibinfo {title} {A
  spectroscopic study of the cis/trans-isomers of penta-2,4-dienoic acid
  attached to gold nanoclusters},}\ }\href {https://doi.org/10.1039/C4CP05280C}
  {\bibfield  {journal} {\bibinfo  {journal} {Physical Chemistry Chemical
  Physics}\ }\textbf {\bibinfo {volume} {17}},\ \bibinfo {pages} {7648--7658}
  (\bibinfo {year} {2015})}\BibitemShut {NoStop}%
\bibitem [{\citenamefont {Barbry}\ \emph {et~al.}(2015)\citenamefont {Barbry},
  \citenamefont {Koval}, \citenamefont {Marchesin}, \citenamefont {Esteban},
  \citenamefont {Borisov}, \citenamefont {Aizpurua},\ and\ \citenamefont
  {{S{\'a}nchez-Portal}}}]{Barbry_NanoLett._2015}%
  \BibitemOpen
  \bibfield  {author} {\bibinfo {author} {\bibfnamefont {M.}~\bibnamefont
  {Barbry}}, \bibinfo {author} {\bibfnamefont {P.}~\bibnamefont {Koval}},
  \bibinfo {author} {\bibfnamefont {F.}~\bibnamefont {Marchesin}}, \bibinfo
  {author} {\bibfnamefont {R.}~\bibnamefont {Esteban}}, \bibinfo {author}
  {\bibfnamefont {A.~G.}\ \bibnamefont {Borisov}}, \bibinfo {author}
  {\bibfnamefont {J.}~\bibnamefont {Aizpurua}},\ and\ \bibinfo {author}
  {\bibfnamefont {D.}~\bibnamefont {{S{\'a}nchez-Portal}}},\ }\bibfield
  {title} {\enquote {\bibinfo {title} {Atomistic {{Near-Field Nanoplasmonics}}:
  {{Reaching Atomic-Scale Resolution}} in {{Nanooptics}}},}\ }\href
  {https://doi.org/10.1021/acs.nanolett.5b00759} {\bibfield  {journal}
  {\bibinfo  {journal} {Nano Letters}\ }\textbf {\bibinfo {volume} {15}},\
  \bibinfo {pages} {3410--3419} (\bibinfo {year} {2015})}\BibitemShut {NoStop}%
\bibitem [{\citenamefont {Schmidt}\ \emph {et~al.}(2016)\citenamefont
  {Schmidt}, \citenamefont {Esteban}, \citenamefont {{Gonz{\'a}lez-Tudela}},
  \citenamefont {Giedke},\ and\ \citenamefont
  {Aizpurua}}]{Schmidt_ACSNano_2016}%
  \BibitemOpen
  \bibfield  {author} {\bibinfo {author} {\bibfnamefont {M.~K.}\ \bibnamefont
  {Schmidt}}, \bibinfo {author} {\bibfnamefont {R.}~\bibnamefont {Esteban}},
  \bibinfo {author} {\bibfnamefont {A.}~\bibnamefont {{Gonz{\'a}lez-Tudela}}},
  \bibinfo {author} {\bibfnamefont {G.}~\bibnamefont {Giedke}},\ and\ \bibinfo
  {author} {\bibfnamefont {J.}~\bibnamefont {Aizpurua}},\ }\bibfield  {title}
  {\enquote {\bibinfo {title} {Quantum {{Mechanical Description}} of {{Raman
  Scattering}} from {{Molecules}} in {{Plasmonic Cavities}}},}\ }\href
  {https://doi.org/10.1021/acsnano.6b02484} {\bibfield  {journal} {\bibinfo
  {journal} {ACS Nano}\ }\textbf {\bibinfo {volume} {10}},\ \bibinfo {pages}
  {6291--6298} (\bibinfo {year} {2016})}\BibitemShut {NoStop}%
\bibitem [{\citenamefont {Zhang}, \citenamefont {Dong},\ and\ \citenamefont
  {Aizpurua}(2021)}]{Zhang_JRamanSpectrosc_2021}%
  \BibitemOpen
  \bibfield  {author} {\bibinfo {author} {\bibfnamefont {Y.}~\bibnamefont
  {Zhang}}, \bibinfo {author} {\bibfnamefont {Z.-C.}\ \bibnamefont {Dong}},\
  and\ \bibinfo {author} {\bibfnamefont {J.}~\bibnamefont {Aizpurua}},\
  }\bibfield  {title} {\enquote {\bibinfo {title} {Theoretical treatment of
  single-molecule scanning {{Raman}} picoscopy in strongly inhomogeneous near
  fields},}\ }\href {https://doi.org/10.1002/jrs.5991} {\bibfield  {journal}
  {\bibinfo  {journal} {Journal of Raman Spectroscopy}\ }\textbf {\bibinfo
  {volume} {52}},\ \bibinfo {pages} {296--309} (\bibinfo {year}
  {2021})}\BibitemShut {NoStop}%
\bibitem [{\citenamefont {Payton}\ \emph {et~al.}(2012)\citenamefont {Payton},
  \citenamefont {Morton}, \citenamefont {Moore},\ and\ \citenamefont
  {Jensen}}]{Payton_J.Chem.Phys._2012}%
  \BibitemOpen
  \bibfield  {author} {\bibinfo {author} {\bibfnamefont {J.~L.}\ \bibnamefont
  {Payton}}, \bibinfo {author} {\bibfnamefont {S.~M.}\ \bibnamefont {Morton}},
  \bibinfo {author} {\bibfnamefont {J.~E.}\ \bibnamefont {Moore}},\ and\
  \bibinfo {author} {\bibfnamefont {L.}~\bibnamefont {Jensen}},\ }\bibfield
  {title} {\enquote {\bibinfo {title} {A discrete interaction model/quantum
  mechanical method for simulating surface-enhanced {{Raman}} spectroscopy},}\
  }\href {https://doi.org/10.1063/1.4722755} {\bibfield  {journal} {\bibinfo
  {journal} {The Journal of Chemical Physics}\ }\textbf {\bibinfo {volume}
  {136}},\ \bibinfo {pages} {214103} (\bibinfo {year} {2012})}\BibitemShut
  {NoStop}%
\bibitem [{\citenamefont {Payton}\ \emph {et~al.}(2014)\citenamefont {Payton},
  \citenamefont {Morton}, \citenamefont {Moore},\ and\ \citenamefont
  {Jensen}}]{Payton_Acc.Chem.Res._2014}%
  \BibitemOpen
  \bibfield  {author} {\bibinfo {author} {\bibfnamefont {J.~L.}\ \bibnamefont
  {Payton}}, \bibinfo {author} {\bibfnamefont {S.~M.}\ \bibnamefont {Morton}},
  \bibinfo {author} {\bibfnamefont {J.~E.}\ \bibnamefont {Moore}},\ and\
  \bibinfo {author} {\bibfnamefont {L.}~\bibnamefont {Jensen}},\ }\bibfield
  {title} {\enquote {\bibinfo {title} {A {{Hybrid Atomistic
  Electrodynamics}}\textendash{{Quantum Mechanical Approach}} for {{Simulating
  Surface-Enhanced Raman Scattering}}},}\ }\href
  {https://doi.org/10.1021/ar400075r} {\bibfield  {journal} {\bibinfo
  {journal} {Accounts of Chemical Research}\ }\textbf {\bibinfo {volume}
  {47}},\ \bibinfo {pages} {88--99} (\bibinfo {year} {2014})}\BibitemShut
  {NoStop}%
\bibitem [{\citenamefont {Hu}, \citenamefont {Chulhai},\ and\ \citenamefont
  {Jensen}(2016)}]{Hu_J.Chem.TheoryComput._2016}%
  \BibitemOpen
  \bibfield  {author} {\bibinfo {author} {\bibfnamefont {Z.}~\bibnamefont
  {Hu}}, \bibinfo {author} {\bibfnamefont {D.~V.}\ \bibnamefont {Chulhai}},\
  and\ \bibinfo {author} {\bibfnamefont {L.}~\bibnamefont {Jensen}},\
  }\bibfield  {title} {\enquote {\bibinfo {title} {Simulating
  {{Surface-Enhanced Hyper-Raman Scattering Using Atomistic
  Electrodynamics-Quantum Mechanical Models}}},}\ }\href
  {https://doi.org/10.1021/acs.jctc.6b00940} {\bibfield  {journal} {\bibinfo
  {journal} {Journal of Chemical Theory and Computation}\ }\textbf {\bibinfo
  {volume} {12}},\ \bibinfo {pages} {5968--5978} (\bibinfo {year}
  {2016})}\BibitemShut {NoStop}%
\bibitem [{\citenamefont {Liu}, \citenamefont {Chulhai},\ and\ \citenamefont
  {Jensen}(2017)}]{Liu_ACSNano_2017}%
  \BibitemOpen
  \bibfield  {author} {\bibinfo {author} {\bibfnamefont {P.}~\bibnamefont
  {Liu}}, \bibinfo {author} {\bibfnamefont {D.~V.}\ \bibnamefont {Chulhai}},\
  and\ \bibinfo {author} {\bibfnamefont {L.}~\bibnamefont {Jensen}},\
  }\bibfield  {title} {\enquote {\bibinfo {title} {Single-{{Molecule Imaging
  Using Atomistic Near-Field Tip-Enhanced Raman Spectroscopy}}},}\ }\href
  {https://doi.org/10.1021/acsnano.7b02058} {\bibfield  {journal} {\bibinfo
  {journal} {ACS Nano}\ }\textbf {\bibinfo {volume} {11}},\ \bibinfo {pages}
  {5094--5102} (\bibinfo {year} {2017})}\BibitemShut {NoStop}%
\bibitem [{\citenamefont {Chen}\ \emph {et~al.}(2019)\citenamefont {Chen},
  \citenamefont {Liu}, \citenamefont {Hu},\ and\ \citenamefont
  {Jensen}}]{Chen_Nat.Commun._2019}%
  \BibitemOpen
  \bibfield  {author} {\bibinfo {author} {\bibfnamefont {X.}~\bibnamefont
  {Chen}}, \bibinfo {author} {\bibfnamefont {P.}~\bibnamefont {Liu}}, \bibinfo
  {author} {\bibfnamefont {Z.}~\bibnamefont {Hu}},\ and\ \bibinfo {author}
  {\bibfnamefont {L.}~\bibnamefont {Jensen}},\ }\bibfield  {title} {\enquote
  {\bibinfo {title} {High-resolution tip-enhanced {{Raman}} scattering probes
  sub-molecular density changes},}\ }\href
  {https://doi.org/10.1038/s41467-019-10618-x} {\bibfield  {journal} {\bibinfo
  {journal} {Nature Communications}\ }\textbf {\bibinfo {volume} {10}},\
  \bibinfo {pages} {2567} (\bibinfo {year} {2019})}\BibitemShut {NoStop}%
\bibitem [{\citenamefont {Hao}\ and\ \citenamefont
  {Schatz}(2003)}]{Hao_J.Chem.Phys._2003}%
  \BibitemOpen
  \bibfield  {author} {\bibinfo {author} {\bibfnamefont {E.}~\bibnamefont
  {Hao}}\ and\ \bibinfo {author} {\bibfnamefont {G.~C.}\ \bibnamefont
  {Schatz}},\ }\bibfield  {title} {\enquote {\bibinfo {title} {Electromagnetic
  fields around silver nanoparticles and dimers},}\ }\href
  {https://doi.org/10.1063/1.1629280} {\bibfield  {journal} {\bibinfo
  {journal} {The Journal of Chemical Physics}\ }\textbf {\bibinfo {volume}
  {120}},\ \bibinfo {pages} {357--366} (\bibinfo {year} {2003})}\BibitemShut
  {NoStop}%
\bibitem [{\citenamefont {Zou}, \citenamefont {Janel},\ and\ \citenamefont
  {Schatz}(2004)}]{Zou_J.Chem.Phys._2004}%
  \BibitemOpen
  \bibfield  {author} {\bibinfo {author} {\bibfnamefont {S.}~\bibnamefont
  {Zou}}, \bibinfo {author} {\bibfnamefont {N.}~\bibnamefont {Janel}},\ and\
  \bibinfo {author} {\bibfnamefont {G.~C.}\ \bibnamefont {Schatz}},\ }\bibfield
   {title} {\enquote {\bibinfo {title} {Silver nanoparticle array structures
  that produce remarkably narrow plasmon lineshapes},}\ }\href
  {https://doi.org/10.1063/1.1760740} {\bibfield  {journal} {\bibinfo
  {journal} {The Journal of Chemical Physics}\ }\textbf {\bibinfo {volume}
  {120}},\ \bibinfo {pages} {10871--10875} (\bibinfo {year}
  {2004})}\BibitemShut {NoStop}%
\bibitem [{\citenamefont {Gieseking}, \citenamefont {Ratner},\ and\
  \citenamefont {Schatz}(2016)}]{Gieseking_J.Phys.Chem.A_2016}%
  \BibitemOpen
  \bibfield  {author} {\bibinfo {author} {\bibfnamefont {R.~L.}\ \bibnamefont
  {Gieseking}}, \bibinfo {author} {\bibfnamefont {M.~A.}\ \bibnamefont
  {Ratner}},\ and\ \bibinfo {author} {\bibfnamefont {G.~C.}\ \bibnamefont
  {Schatz}},\ }\bibfield  {title} {\enquote {\bibinfo {title} {Semiempirical
  {{Modeling}} of {{Ag Nanoclusters}}: {{New Parameters}} for {{Optical
  Property Studies Enable Determination}} of {{Double Excitation
  Contributions}} to {{Plasmonic Excitation}}},}\ }\href
  {https://doi.org/10.1021/acs.jpca.6b04520} {\bibfield  {journal} {\bibinfo
  {journal} {The Journal of Physical Chemistry A}\ }\textbf {\bibinfo {volume}
  {120}},\ \bibinfo {pages} {4542--4549} (\bibinfo {year} {2016})}\BibitemShut
  {NoStop}%
\bibitem [{\citenamefont {Ding}\ \emph {et~al.}(2018)\citenamefont {Ding},
  \citenamefont {Hsu}, \citenamefont {Heaps},\ and\ \citenamefont
  {Schatz}}]{Ding_J.Phys.Chem.C_2018}%
  \BibitemOpen
  \bibfield  {author} {\bibinfo {author} {\bibfnamefont {W.}~\bibnamefont
  {Ding}}, \bibinfo {author} {\bibfnamefont {L.-Y.}\ \bibnamefont {Hsu}},
  \bibinfo {author} {\bibfnamefont {C.~W.}\ \bibnamefont {Heaps}},\ and\
  \bibinfo {author} {\bibfnamefont {G.~C.}\ \bibnamefont {Schatz}},\ }\bibfield
   {title} {\enquote {\bibinfo {title} {Plasmon-{{Coupled Resonance Energy
  Transfer II}}: {{Exploring}} the {{Peaks}} and {{Dips}} in the
  {{Electromagnetic Coupling Factor}}},}\ }\href
  {https://doi.org/10.1021/acs.jpcc.8b07210} {\bibfield  {journal} {\bibinfo
  {journal} {The Journal of Physical Chemistry C}\ }\textbf {\bibinfo {volume}
  {122}},\ \bibinfo {pages} {22650--22659} (\bibinfo {year}
  {2018})}\BibitemShut {NoStop}%
\bibitem [{\citenamefont {Kluender}\ \emph {et~al.}(2021)\citenamefont
  {Kluender}, \citenamefont {Bourgeois}, \citenamefont {Cherqui}, \citenamefont
  {Schatz},\ and\ \citenamefont {Mirkin}}]{Kluender_J.Phys.Chem.C_2021}%
  \BibitemOpen
  \bibfield  {author} {\bibinfo {author} {\bibfnamefont {E.~J.}\ \bibnamefont
  {Kluender}}, \bibinfo {author} {\bibfnamefont {M.~R.}\ \bibnamefont
  {Bourgeois}}, \bibinfo {author} {\bibfnamefont {C.~R.}\ \bibnamefont
  {Cherqui}}, \bibinfo {author} {\bibfnamefont {G.~C.}\ \bibnamefont
  {Schatz}},\ and\ \bibinfo {author} {\bibfnamefont {C.~A.}\ \bibnamefont
  {Mirkin}},\ }\bibfield  {title} {\enquote {\bibinfo {title} {Multimetallic
  {{Nanoparticles}} on {{Mirrors}} for {{SERS Detection}}},}\ }\href
  {https://doi.org/10.1021/acs.jpcc.1c02578} {\bibfield  {journal} {\bibinfo
  {journal} {The Journal of Physical Chemistry C}\ }\textbf {\bibinfo {volume}
  {125}},\ \bibinfo {pages} {12784--12791} (\bibinfo {year}
  {2021})}\BibitemShut {NoStop}%
\bibitem [{\citenamefont {Morton}\ and\ \citenamefont
  {Jensen}(2010)}]{Morton_J.Chem.Phys._2010}%
  \BibitemOpen
  \bibfield  {author} {\bibinfo {author} {\bibfnamefont {S.~M.}\ \bibnamefont
  {Morton}}\ and\ \bibinfo {author} {\bibfnamefont {L.}~\bibnamefont
  {Jensen}},\ }\bibfield  {title} {\enquote {\bibinfo {title} {A discrete
  interaction model/quantum mechanical method for describing response
  properties of molecules adsorbed on metal nanoparticles},}\ }\href
  {https://doi.org/10.1063/1.3457365} {\bibfield  {journal} {\bibinfo
  {journal} {The Journal of Chemical Physics}\ }\textbf {\bibinfo {volume}
  {133}},\ \bibinfo {pages} {074103} (\bibinfo {year} {2010})}\BibitemShut
  {NoStop}%
\bibitem [{\citenamefont {Morton}\ and\ \citenamefont
  {Jensen}(2011)}]{Morton_J.Chem.Phys._2011}%
  \BibitemOpen
  \bibfield  {author} {\bibinfo {author} {\bibfnamefont {S.~M.}\ \bibnamefont
  {Morton}}\ and\ \bibinfo {author} {\bibfnamefont {L.}~\bibnamefont
  {Jensen}},\ }\bibfield  {title} {\enquote {\bibinfo {title} {A discrete
  interaction model/quantum mechanical method to describe the interaction of
  metal nanoparticles and molecular absorption},}\ }\href
  {https://doi.org/10.1063/1.3643381} {\bibfield  {journal} {\bibinfo
  {journal} {The Journal of Chemical Physics}\ }\textbf {\bibinfo {volume}
  {135}},\ \bibinfo {pages} {134103} (\bibinfo {year} {2011})}\BibitemShut
  {NoStop}%
\bibitem [{\citenamefont {Zhao}\ \emph {et~al.}(2008)\citenamefont {Zhao},
  \citenamefont {Pinchuk}, \citenamefont {McMahon}, \citenamefont {Li},
  \citenamefont {Ausman}, \citenamefont {Atkinson},\ and\ \citenamefont
  {Schatz}}]{Zhao_Acc.Chem.Res._2008}%
  \BibitemOpen
  \bibfield  {author} {\bibinfo {author} {\bibfnamefont {J.}~\bibnamefont
  {Zhao}}, \bibinfo {author} {\bibfnamefont {A.~O.}\ \bibnamefont {Pinchuk}},
  \bibinfo {author} {\bibfnamefont {J.~M.}\ \bibnamefont {McMahon}}, \bibinfo
  {author} {\bibfnamefont {S.}~\bibnamefont {Li}}, \bibinfo {author}
  {\bibfnamefont {L.~K.}\ \bibnamefont {Ausman}}, \bibinfo {author}
  {\bibfnamefont {A.~L.}\ \bibnamefont {Atkinson}},\ and\ \bibinfo {author}
  {\bibfnamefont {G.~C.}\ \bibnamefont {Schatz}},\ }\bibfield  {title}
  {\enquote {\bibinfo {title} {Methods for {{Describing}} the {{Electromagnetic
  Properties}} of {{Silver}} and {{Gold Nanoparticles}}},}\ }\href
  {https://doi.org/10.1021/ar800028j} {\bibfield  {journal} {\bibinfo
  {journal} {Accounts of Chemical Research}\ }\textbf {\bibinfo {volume}
  {41}},\ \bibinfo {pages} {1710--1720} (\bibinfo {year} {2008})}\BibitemShut
  {NoStop}%
\bibitem [{\citenamefont {Latorre}\ \emph {et~al.}(2016)\citenamefont
  {Latorre}, \citenamefont {Kupfer}, \citenamefont {Bocklitz}, \citenamefont
  {Kinzel}, \citenamefont {Trautmann}, \citenamefont {Gr{\"a}fe},\ and\
  \citenamefont {Deckert}}]{Latorre_Nanoscale_2016}%
  \BibitemOpen
  \bibfield  {author} {\bibinfo {author} {\bibfnamefont {F.}~\bibnamefont
  {Latorre}}, \bibinfo {author} {\bibfnamefont {S.}~\bibnamefont {Kupfer}},
  \bibinfo {author} {\bibfnamefont {T.}~\bibnamefont {Bocklitz}}, \bibinfo
  {author} {\bibfnamefont {D.}~\bibnamefont {Kinzel}}, \bibinfo {author}
  {\bibfnamefont {S.}~\bibnamefont {Trautmann}}, \bibinfo {author}
  {\bibfnamefont {S.}~\bibnamefont {Gr{\"a}fe}},\ and\ \bibinfo {author}
  {\bibfnamefont {V.}~\bibnamefont {Deckert}},\ }\bibfield  {title} {\enquote
  {\bibinfo {title} {Spatial resolution of tip-enhanced {{Raman}} spectroscopy
  \textendash{} {{DFT}} assessment of the chemical effect},}\ }\href
  {https://doi.org/10.1039/C6NR00093B} {\bibfield  {journal} {\bibinfo
  {journal} {Nanoscale}\ }\textbf {\bibinfo {volume} {8}},\ \bibinfo {pages}
  {10229--10239} (\bibinfo {year} {2016})}\BibitemShut {NoStop}%
\bibitem [{\citenamefont {Fiederling}\ \emph {et~al.}(2020)\citenamefont
  {Fiederling}, \citenamefont {Abasifard}, \citenamefont {Richter},
  \citenamefont {Deckert}, \citenamefont {Gr{\"a}fe},\ and\ \citenamefont
  {Kupfer}}]{Fiederling_Nanoscale_2020}%
  \BibitemOpen
  \bibfield  {author} {\bibinfo {author} {\bibfnamefont {K.}~\bibnamefont
  {Fiederling}}, \bibinfo {author} {\bibfnamefont {M.}~\bibnamefont
  {Abasifard}}, \bibinfo {author} {\bibfnamefont {M.}~\bibnamefont {Richter}},
  \bibinfo {author} {\bibfnamefont {V.}~\bibnamefont {Deckert}}, \bibinfo
  {author} {\bibfnamefont {S.}~\bibnamefont {Gr{\"a}fe}},\ and\ \bibinfo
  {author} {\bibfnamefont {S.}~\bibnamefont {Kupfer}},\ }\bibfield  {title}
  {\enquote {\bibinfo {title} {The chemical effect goes resonant \textendash{}
  a full quantum mechanical approach on {{TERS}}},}\ }\href
  {https://doi.org/10.1039/C9NR09814C} {\bibfield  {journal} {\bibinfo
  {journal} {Nanoscale}\ }\textbf {\bibinfo {volume} {12}},\ \bibinfo {pages}
  {6346--6359} (\bibinfo {year} {2020})}\BibitemShut {NoStop}%
\bibitem [{\citenamefont {Fiederling}, \citenamefont {Kupfer},\ and\
  \citenamefont {Gr{\"a}fe}(2021)}]{Fiederling_J.Chem.Phys._2021}%
  \BibitemOpen
  \bibfield  {author} {\bibinfo {author} {\bibfnamefont {K.}~\bibnamefont
  {Fiederling}}, \bibinfo {author} {\bibfnamefont {S.}~\bibnamefont {Kupfer}},\
  and\ \bibinfo {author} {\bibfnamefont {S.}~\bibnamefont {Gr{\"a}fe}},\
  }\bibfield  {title} {\enquote {\bibinfo {title} {Are charged tips driving
  {{TERS-resolution}}? {{A}} full quantum chemical approach},}\ }\href
  {https://doi.org/10.1063/5.0031763} {\bibfield  {journal} {\bibinfo
  {journal} {The Journal of Chemical Physics}\ }\textbf {\bibinfo {volume}
  {154}},\ \bibinfo {pages} {034106} (\bibinfo {year} {2021})}\BibitemShut
  {NoStop}%
\bibitem [{\citenamefont {Rodriguez}\ \emph {et~al.}(2021)\citenamefont
  {Rodriguez}, \citenamefont {Villag{\'o}mez}, \citenamefont {Khodadadi},
  \citenamefont {Kupfer}, \citenamefont {Averkiev}, \citenamefont {Dedelaite},
  \citenamefont {Tang}, \citenamefont {Khaywah}, \citenamefont {Kolchuzhin},
  \citenamefont {Ramanavicius}, \citenamefont {Adam}, \citenamefont
  {Gr{\"a}fe},\ and\ \citenamefont {Sheremet}}]{Rodriguez_ACSPhotonics_2021}%
  \BibitemOpen
  \bibfield  {author} {\bibinfo {author} {\bibfnamefont {R.~D.}\ \bibnamefont
  {Rodriguez}}, \bibinfo {author} {\bibfnamefont {C.~J.}\ \bibnamefont
  {Villag{\'o}mez}}, \bibinfo {author} {\bibfnamefont {A.}~\bibnamefont
  {Khodadadi}}, \bibinfo {author} {\bibfnamefont {S.}~\bibnamefont {Kupfer}},
  \bibinfo {author} {\bibfnamefont {A.}~\bibnamefont {Averkiev}}, \bibinfo
  {author} {\bibfnamefont {L.}~\bibnamefont {Dedelaite}}, \bibinfo {author}
  {\bibfnamefont {F.}~\bibnamefont {Tang}}, \bibinfo {author} {\bibfnamefont
  {M.~Y.}\ \bibnamefont {Khaywah}}, \bibinfo {author} {\bibfnamefont
  {V.}~\bibnamefont {Kolchuzhin}}, \bibinfo {author} {\bibfnamefont
  {A.}~\bibnamefont {Ramanavicius}}, \bibinfo {author} {\bibfnamefont {P.-M.}\
  \bibnamefont {Adam}}, \bibinfo {author} {\bibfnamefont {S.}~\bibnamefont
  {Gr{\"a}fe}},\ and\ \bibinfo {author} {\bibfnamefont {E.}~\bibnamefont
  {Sheremet}},\ }\bibfield  {title} {\enquote {\bibinfo {title} {Chemical
  {{Enhancement}} vs {{Molecule}}\textendash{{Substrate Geometry}} in
  {{Plasmon-Enhanced Spectroscopy}}},}\ }\href
  {https://doi.org/10.1021/acsphotonics.1c00001} {\bibfield  {journal}
  {\bibinfo  {journal} {ACS Photonics}\ }\textbf {\bibinfo {volume} {8}},\
  \bibinfo {pages} {2243--2255} (\bibinfo {year} {2021})}\BibitemShut {NoStop}%
\bibitem [{com()}]{comsol}%
  \BibitemOpen
  \href@noop {} {\enquote {\bibinfo {title} {{{COMSOL Multiphysics}}},}\
  }\bibinfo {howpublished} {COMSOL AB}\BibitemShut {NoStop}%
\bibitem [{\citenamefont {Johnson}\ and\ \citenamefont
  {Christy}(1972)}]{Johnson_Phys.Rev.B_1972}%
  \BibitemOpen
  \bibfield  {author} {\bibinfo {author} {\bibfnamefont {P.~B.}\ \bibnamefont
  {Johnson}}\ and\ \bibinfo {author} {\bibfnamefont {R.~W.}\ \bibnamefont
  {Christy}},\ }\bibfield  {title} {\enquote {\bibinfo {title} {Optical
  {{Constants}} of the {{Noble Metals}}},}\ }\href
  {https://doi.org/10.1103/PhysRevB.6.4370} {\bibfield  {journal} {\bibinfo
  {journal} {Physical Review B}\ }\textbf {\bibinfo {volume} {6}},\ \bibinfo
  {pages} {4370--4379} (\bibinfo {year} {1972})}\BibitemShut {NoStop}%
\bibitem [{\citenamefont {Sakko}, \citenamefont {Rossi},\ and\ \citenamefont
  {Nieminen}(2014)}]{Sakko_JPhysCondensMatter_2014}%
  \BibitemOpen
  \bibfield  {author} {\bibinfo {author} {\bibfnamefont {A.}~\bibnamefont
  {Sakko}}, \bibinfo {author} {\bibfnamefont {T.~P.}\ \bibnamefont {Rossi}},\
  and\ \bibinfo {author} {\bibfnamefont {R.~M.}\ \bibnamefont {Nieminen}},\
  }\bibfield  {title} {\enquote {\bibinfo {title} {Dynamical coupling of
  plasmons and molecular excitations by hybrid quantum/classical calculations:
  Time-domain approach},}\ }\href
  {https://doi.org/10.1088/0953-8984/26/28/315013} {\bibfield  {journal}
  {\bibinfo  {journal} {Journal of Physics. Condensed Matter: An Institute of
  Physics Journal}\ }\textbf {\bibinfo {volume} {26}},\ \bibinfo {pages}
  {315013} (\bibinfo {year} {2014})}\BibitemShut {NoStop}%
\bibitem [{\citenamefont {Frisch}\ \emph {et~al.}(2016)\citenamefont {Frisch},
  \citenamefont {Trucks}, \citenamefont {Schlegel}, \citenamefont {Scuseria},
  \citenamefont {Robb}, \citenamefont {Cheeseman}, \citenamefont {Scalmani},
  \citenamefont {Barone}, \citenamefont {Petersson}, \citenamefont {Nakatsuji},
  \citenamefont {Li}, \citenamefont {Caricato}, \citenamefont {Marenich},
  \citenamefont {Bloino}, \citenamefont {Janesko}, \citenamefont {Gomperts},
  \citenamefont {Mennucci}, \citenamefont {Hratchian}, \citenamefont {Ortiz},
  \citenamefont {Izmaylov}, \citenamefont {Sonnenberg}, \citenamefont
  {{Williams-Young}}, \citenamefont {Ding}, \citenamefont {Lipparini},
  \citenamefont {Egidi}, \citenamefont {Goings}, \citenamefont {Peng},
  \citenamefont {Petrone}, \citenamefont {Henderson}, \citenamefont
  {Ranasinghe}, \citenamefont {Zakrzewski}, \citenamefont {Gao}, \citenamefont
  {Rega}, \citenamefont {Zheng}, \citenamefont {Liang}, \citenamefont {Hada},
  \citenamefont {Ehara}, \citenamefont {Toyota}, \citenamefont {Fukuda},
  \citenamefont {Hasegawa}, \citenamefont {Ishida}, \citenamefont {Nakajima},
  \citenamefont {Honda}, \citenamefont {Kitao}, \citenamefont {Nakai},
  \citenamefont {Vreven}, \citenamefont {Throssell}, \citenamefont
  {Montgomery}, \citenamefont {Peralta}, \citenamefont {Ogliaro}, \citenamefont
  {Bearpark}, \citenamefont {Heyd}, \citenamefont {Brothers}, \citenamefont
  {Kudin}, \citenamefont {Staroverov}, \citenamefont {Keith}, \citenamefont
  {Kobayashi}, \citenamefont {Normand}, \citenamefont {Raghavachari},
  \citenamefont {Rendell}, \citenamefont {Burant}, \citenamefont {Iyengar},
  \citenamefont {Tomasi}, \citenamefont {Cossi}, \citenamefont {Millam},
  \citenamefont {Klene}, \citenamefont {Adamo}, \citenamefont {Cammi},
  \citenamefont {Ochterski}, \citenamefont {Martin}, \citenamefont {Morokuma},
  \citenamefont {Farkas}, \citenamefont {Foresman},\ and\ \citenamefont
  {Fox}}]{g16}%
  \BibitemOpen
  \bibfield  {author} {\bibinfo {author} {\bibfnamefont {M.~J.}\ \bibnamefont
  {Frisch}}, \bibinfo {author} {\bibfnamefont {G.~W.}\ \bibnamefont {Trucks}},
  \bibinfo {author} {\bibfnamefont {H.~B.}\ \bibnamefont {Schlegel}}, \bibinfo
  {author} {\bibfnamefont {G.~E.}\ \bibnamefont {Scuseria}}, \bibinfo {author}
  {\bibfnamefont {M.~A.}\ \bibnamefont {Robb}}, \bibinfo {author}
  {\bibfnamefont {J.~R.}\ \bibnamefont {Cheeseman}}, \bibinfo {author}
  {\bibfnamefont {G.}~\bibnamefont {Scalmani}}, \bibinfo {author}
  {\bibfnamefont {V.}~\bibnamefont {Barone}}, \bibinfo {author} {\bibfnamefont
  {G.~A.}\ \bibnamefont {Petersson}}, \bibinfo {author} {\bibfnamefont
  {H.}~\bibnamefont {Nakatsuji}}, \bibinfo {author} {\bibfnamefont
  {X.}~\bibnamefont {Li}}, \bibinfo {author} {\bibfnamefont {M.}~\bibnamefont
  {Caricato}}, \bibinfo {author} {\bibfnamefont {A.~V.}\ \bibnamefont
  {Marenich}}, \bibinfo {author} {\bibfnamefont {J.}~\bibnamefont {Bloino}},
  \bibinfo {author} {\bibfnamefont {B.~G.}\ \bibnamefont {Janesko}}, \bibinfo
  {author} {\bibfnamefont {R.}~\bibnamefont {Gomperts}}, \bibinfo {author}
  {\bibfnamefont {B.}~\bibnamefont {Mennucci}}, \bibinfo {author}
  {\bibfnamefont {H.~P.}\ \bibnamefont {Hratchian}}, \bibinfo {author}
  {\bibfnamefont {J.~V.}\ \bibnamefont {Ortiz}}, \bibinfo {author}
  {\bibfnamefont {A.~F.}\ \bibnamefont {Izmaylov}}, \bibinfo {author}
  {\bibfnamefont {J.~L.}\ \bibnamefont {Sonnenberg}}, \bibinfo {author}
  {\bibfnamefont {D.}~\bibnamefont {{Williams-Young}}}, \bibinfo {author}
  {\bibfnamefont {F.}~\bibnamefont {Ding}}, \bibinfo {author} {\bibfnamefont
  {F.}~\bibnamefont {Lipparini}}, \bibinfo {author} {\bibfnamefont
  {F.}~\bibnamefont {Egidi}}, \bibinfo {author} {\bibfnamefont
  {J.}~\bibnamefont {Goings}}, \bibinfo {author} {\bibfnamefont
  {B.}~\bibnamefont {Peng}}, \bibinfo {author} {\bibfnamefont {A.}~\bibnamefont
  {Petrone}}, \bibinfo {author} {\bibfnamefont {T.}~\bibnamefont {Henderson}},
  \bibinfo {author} {\bibfnamefont {D.}~\bibnamefont {Ranasinghe}}, \bibinfo
  {author} {\bibfnamefont {V.~G.}\ \bibnamefont {Zakrzewski}}, \bibinfo
  {author} {\bibfnamefont {J.}~\bibnamefont {Gao}}, \bibinfo {author}
  {\bibfnamefont {N.}~\bibnamefont {Rega}}, \bibinfo {author} {\bibfnamefont
  {G.}~\bibnamefont {Zheng}}, \bibinfo {author} {\bibfnamefont
  {W.}~\bibnamefont {Liang}}, \bibinfo {author} {\bibfnamefont
  {M.}~\bibnamefont {Hada}}, \bibinfo {author} {\bibfnamefont {M.}~\bibnamefont
  {Ehara}}, \bibinfo {author} {\bibfnamefont {K.}~\bibnamefont {Toyota}},
  \bibinfo {author} {\bibfnamefont {R.}~\bibnamefont {Fukuda}}, \bibinfo
  {author} {\bibfnamefont {J.}~\bibnamefont {Hasegawa}}, \bibinfo {author}
  {\bibfnamefont {M.}~\bibnamefont {Ishida}}, \bibinfo {author} {\bibfnamefont
  {T.}~\bibnamefont {Nakajima}}, \bibinfo {author} {\bibfnamefont
  {Y.}~\bibnamefont {Honda}}, \bibinfo {author} {\bibfnamefont
  {O.}~\bibnamefont {Kitao}}, \bibinfo {author} {\bibfnamefont
  {H.}~\bibnamefont {Nakai}}, \bibinfo {author} {\bibfnamefont
  {T.}~\bibnamefont {Vreven}}, \bibinfo {author} {\bibfnamefont
  {K.}~\bibnamefont {Throssell}}, \bibinfo {author} {\bibfnamefont {J.~A.}\
  \bibnamefont {Montgomery}, \bibfnamefont {Jr.}}, \bibinfo {author}
  {\bibfnamefont {J.~E.}\ \bibnamefont {Peralta}}, \bibinfo {author}
  {\bibfnamefont {F.}~\bibnamefont {Ogliaro}}, \bibinfo {author} {\bibfnamefont
  {M.~J.}\ \bibnamefont {Bearpark}}, \bibinfo {author} {\bibfnamefont {J.~J.}\
  \bibnamefont {Heyd}}, \bibinfo {author} {\bibfnamefont {E.~N.}\ \bibnamefont
  {Brothers}}, \bibinfo {author} {\bibfnamefont {K.~N.}\ \bibnamefont {Kudin}},
  \bibinfo {author} {\bibfnamefont {V.~N.}\ \bibnamefont {Staroverov}},
  \bibinfo {author} {\bibfnamefont {T.~A.}\ \bibnamefont {Keith}}, \bibinfo
  {author} {\bibfnamefont {R.}~\bibnamefont {Kobayashi}}, \bibinfo {author}
  {\bibfnamefont {J.}~\bibnamefont {Normand}}, \bibinfo {author} {\bibfnamefont
  {K.}~\bibnamefont {Raghavachari}}, \bibinfo {author} {\bibfnamefont {A.~P.}\
  \bibnamefont {Rendell}}, \bibinfo {author} {\bibfnamefont {J.~C.}\
  \bibnamefont {Burant}}, \bibinfo {author} {\bibfnamefont {S.~S.}\
  \bibnamefont {Iyengar}}, \bibinfo {author} {\bibfnamefont {J.}~\bibnamefont
  {Tomasi}}, \bibinfo {author} {\bibfnamefont {M.}~\bibnamefont {Cossi}},
  \bibinfo {author} {\bibfnamefont {J.~M.}\ \bibnamefont {Millam}}, \bibinfo
  {author} {\bibfnamefont {M.}~\bibnamefont {Klene}}, \bibinfo {author}
  {\bibfnamefont {C.}~\bibnamefont {Adamo}}, \bibinfo {author} {\bibfnamefont
  {R.}~\bibnamefont {Cammi}}, \bibinfo {author} {\bibfnamefont {J.~W.}\
  \bibnamefont {Ochterski}}, \bibinfo {author} {\bibfnamefont {R.~L.}\
  \bibnamefont {Martin}}, \bibinfo {author} {\bibfnamefont {K.}~\bibnamefont
  {Morokuma}}, \bibinfo {author} {\bibfnamefont {O.}~\bibnamefont {Farkas}},
  \bibinfo {author} {\bibfnamefont {J.~B.}\ \bibnamefont {Foresman}},\ and\
  \bibinfo {author} {\bibfnamefont {D.~J.}\ \bibnamefont {Fox}},\ }\href@noop
  {} {\enquote {\bibinfo {title} {Gaussian 16 {{Revision B}}.01},}\ } (\bibinfo
  {year} {2016})\BibitemShut {NoStop}%
\bibitem [{\citenamefont {Yanai}, \citenamefont {Tew},\ and\ \citenamefont
  {Handy}(2004)}]{Yanai_ChemicalPhysicsLetters_2004}%
  \BibitemOpen
  \bibfield  {author} {\bibinfo {author} {\bibfnamefont {T.}~\bibnamefont
  {Yanai}}, \bibinfo {author} {\bibfnamefont {D.~P.}\ \bibnamefont {Tew}},\
  and\ \bibinfo {author} {\bibfnamefont {N.~C.}\ \bibnamefont {Handy}},\
  }\bibfield  {title} {\enquote {\bibinfo {title} {A new hybrid
  exchange\textendash correlation functional using the {{Coulomb-attenuating}}
  method ({{CAM-B3LYP}})},}\ }\href
  {https://doi.org/10.1016/j.cplett.2004.06.011} {\bibfield  {journal}
  {\bibinfo  {journal} {Chemical Physics Letters}\ }\textbf {\bibinfo {volume}
  {393}},\ \bibinfo {pages} {51--57} (\bibinfo {year} {2004})}\BibitemShut
  {NoStop}%
\bibitem [{\citenamefont {Krishnan}\ \emph {et~al.}(1980)\citenamefont
  {Krishnan}, \citenamefont {Binkley}, \citenamefont {Seeger},\ and\
  \citenamefont {Pople}}]{Krishnan_J.Chem.Phys._1980}%
  \BibitemOpen
  \bibfield  {author} {\bibinfo {author} {\bibfnamefont {R.}~\bibnamefont
  {Krishnan}}, \bibinfo {author} {\bibfnamefont {J.~S.}\ \bibnamefont
  {Binkley}}, \bibinfo {author} {\bibfnamefont {R.}~\bibnamefont {Seeger}},\
  and\ \bibinfo {author} {\bibfnamefont {J.~A.}\ \bibnamefont {Pople}},\
  }\bibfield  {title} {\enquote {\bibinfo {title} {Self-consistent molecular
  orbital methods. {{XX}}. {{A}} basis set for correlated wave functions},}\
  }\href {https://doi.org/10.1063/1.438955} {\bibfield  {journal} {\bibinfo
  {journal} {The Journal of Chemical Physics}\ }\textbf {\bibinfo {volume}
  {72}},\ \bibinfo {pages} {650--654} (\bibinfo {year} {1980})}\BibitemShut
  {NoStop}%
\bibitem [{\citenamefont {Clark}\ \emph {et~al.}(1983)\citenamefont {Clark},
  \citenamefont {Chandrasekhar}, \citenamefont {Spitznagel},\ and\
  \citenamefont {Schleyer}}]{Clark_J.Comput.Chem._1983}%
  \BibitemOpen
  \bibfield  {author} {\bibinfo {author} {\bibfnamefont {T.}~\bibnamefont
  {Clark}}, \bibinfo {author} {\bibfnamefont {J.}~\bibnamefont
  {Chandrasekhar}}, \bibinfo {author} {\bibfnamefont {G.~W.}\ \bibnamefont
  {Spitznagel}},\ and\ \bibinfo {author} {\bibfnamefont {P.~V.~R.}\
  \bibnamefont {Schleyer}},\ }\bibfield  {title} {\enquote {\bibinfo {title}
  {Efficient diffuse function-augmented basis sets for anion calculations.
  {{III}}. {{The}} 3-21+{{G}} basis set for first-row elements,
  {{Li}}\textendash{{F}}},}\ }\href {https://doi.org/10.1002/jcc.540040303}
  {\bibfield  {journal} {\bibinfo  {journal} {Journal of Computational
  Chemistry}\ }\textbf {\bibinfo {volume} {4}},\ \bibinfo {pages} {294--301}
  (\bibinfo {year} {1983})}\BibitemShut {NoStop}%
\bibitem [{\citenamefont {Andrae}\ \emph {et~al.}(1990)\citenamefont {Andrae},
  \citenamefont {H{\"a}u{\ss}ermann}, \citenamefont {Dolg}, \citenamefont
  {Stoll},\ and\ \citenamefont {Preu{\ss}}}]{Andrae_Theoret.Chim.Acta_1990}%
  \BibitemOpen
  \bibfield  {author} {\bibinfo {author} {\bibfnamefont {D.}~\bibnamefont
  {Andrae}}, \bibinfo {author} {\bibfnamefont {U.}~\bibnamefont
  {H{\"a}u{\ss}ermann}}, \bibinfo {author} {\bibfnamefont {M.}~\bibnamefont
  {Dolg}}, \bibinfo {author} {\bibfnamefont {H.}~\bibnamefont {Stoll}},\ and\
  \bibinfo {author} {\bibfnamefont {H.}~\bibnamefont {Preu{\ss}}},\ }\bibfield
  {title} {\enquote {\bibinfo {title} {Energy-adjusted ab initio
  pseudopotentials for the second and third row transition elements},}\ }\href
  {https://doi.org/10.1007/BF01114537} {\bibfield  {journal} {\bibinfo
  {journal} {Theoretica chimica acta}\ }\textbf {\bibinfo {volume} {77}},\
  \bibinfo {pages} {123--141} (\bibinfo {year} {1990})}\BibitemShut {NoStop}%
\bibitem [{\citenamefont {Grimme}, \citenamefont {Ehrlich},\ and\ \citenamefont
  {Goerigk}(2011)}]{Grimme_J.Comput.Chem._2011}%
  \BibitemOpen
  \bibfield  {author} {\bibinfo {author} {\bibfnamefont {S.}~\bibnamefont
  {Grimme}}, \bibinfo {author} {\bibfnamefont {S.}~\bibnamefont {Ehrlich}},\
  and\ \bibinfo {author} {\bibfnamefont {L.}~\bibnamefont {Goerigk}},\
  }\bibfield  {title} {\enquote {\bibinfo {title} {Effect of the damping
  function in dispersion corrected density functional theory},}\ }\href
  {https://doi.org/10.1002/jcc.21759} {\bibfield  {journal} {\bibinfo
  {journal} {Journal of Computational Chemistry}\ }\textbf {\bibinfo {volume}
  {32}},\ \bibinfo {pages} {1456--1465} (\bibinfo {year} {2011})}\BibitemShut
  {NoStop}%
\bibitem [{\citenamefont {Guthmuller}(2016)}]{Guthmuller_J.Chem.Phys._2016}%
  \BibitemOpen
  \bibfield  {author} {\bibinfo {author} {\bibfnamefont {J.}~\bibnamefont
  {Guthmuller}},\ }\bibfield  {title} {\enquote {\bibinfo {title} {Comparison
  of simplified sum-over-state expressions to calculate resonance {{Raman}}
  intensities including {{Franck-Condon}} and {{Herzberg-Teller}} effects},}\
  }\href {https://doi.org/10.1063/1.4941449} {\bibfield  {journal} {\bibinfo
  {journal} {The Journal of Chemical Physics}\ }\textbf {\bibinfo {volume}
  {144}},\ \bibinfo {pages} {064106} (\bibinfo {year} {2016})}\BibitemShut
  {NoStop}%
\end{thebibliography}%

\end{document}